\documentclass[twocolumn]{aa}

\usepackage[T1]{fontenc}
\usepackage{ae,aecompl}
\usepackage{lipsum}
\usepackage{lscape}
\usepackage{natbib}

\usepackage{listings}
\usepackage{realboxes}
\bibpunct{(}{)}{;}{a}{}{,} 

\usepackage{graphicx}	
\usepackage{amsmath}	
\usepackage{amssymb}	
\usepackage[dvipsnames]{xcolor}
\usepackage{txfonts}
\usepackage[colorlinks=true,allcolors=cyan]{hyperref}
\usepackage{ulem}
\usepackage{multirow}

\defcitealias{zhang2024hot}{Z24}
\defcitealias{aird2013primus}{A13}
\defcitealias{bekhti2016hi4pi}{HI4PI collaboration et al. 2016}
\defcitealias{ade2016planck}{Planck Collaboration XIII 2016}
\defcitealias{anderson2015unifying}{A15}
\newcommand{\MSUN}{{\rm M}_\odot}

\usepackage{titlesec}

\titleformat{\paragraph}
{\normalfont\normalsize}{\theparagraph}{1em}{}
\titlespacing*{\paragraph}
{0pt}{2.25ex plus 1ex minus .2ex}{1.5ex plus .2ex}

\begin{document} 

   \title{Retrieving the hot circumgalactic medium physics from the X-ray radial profile from eROSITA with an IlustrisTNG-based forward model}

    \author{Soumya Shreeram\inst{1}\thanks{\href{shreeram@mpe.mpg.de}{shreeram@mpe.mpg.de}}, Johan Comparat\inst{1}, Andrea Merloni\inst{1}, Gabriele Ponti\inst{1,2,3}, Paola Popesso\inst{4}, Yi Zhang\inst{1}, Kirpal Nandra\inst{1}, Mara Salvato\inst{1},  Ilaria Marini\inst{4}, Johannes Buchner\inst{1}, Nicola Locatelli\inst{2}, Zsofi Igo\inst{1,5}}

    \institute{Max Planck Institute for Extraterrestrial Physics (MPE), Gie\ss enbachstraße 1, 85748 Garching, Munich, Germany
    \and
        INAF-Osservatorio Astronomico di Brera, Via E. Bianchi 46, I-23807 Merate (LC), Italy
    \and 
        Como Lake Center for Astrophysics (CLAP), DiSAT, Università degli Studi dell'Insubria, via Valleggio 11, I-22100 Como, Italy
    \and
        European Southern Observatory, Karl-Schwarzschild-Stra\ss e 2, 85748 Garching, Munich, Germany
    \and
        Exzellenzcluster ORIGINS, Boltzmannstr. 2, 85748, Garching, Germany
    }
    \date{Received ZZZZ, ZZZZ; accepted XXXX, XXXX; published YYYY, YYYY}

\abstract
  {}
  {Recent eROSITA measurements of the radial profiles of the hot circumgalactic medium (CGM) in the Milky Way stellar mass (MW-mass) regime provide us with a new benchmark to constrain the hot gas around MW-mass central and satellite galaxies and their halo mass distributions. Modeling this rich data set with state-of-the-art hydrodynamical simulations is required to further our understanding of the shortcomings in the current paradigm of galaxy formation and evolution models.}
  {We developed forward models for the stacked X-ray radial surface brightness profile measured by eROSITA around MW-mass galaxies. Our model contains two emitting components: hot gas (around central galaxies and around satellite galaxies hosted by more massive halos) and X-ray point sources (X-ray binaries (XRBs) and active galactic nuclei (AGNs)). We modeled the hot gas profile using the TNG300-based products. We generated mock observations with our TNG300-based model (matching stellar mass and redshift with observations) with different underlying halo mass distributions. Therefore, we tested the CGM properties as a function of their host halo mass distribution. The point sources are described by a simple point spread function of eROSITA, and we fit their normalization in this work. In total, we fit the X-ray surface brightness profile with two free parameters: the normalization of satellites in more massive host halos and the normalization of the mean point source emission.
  }
  {We show that for the same mean stellar mass, a factor $\sim 2\times$ increase in the mean value of the underlying halo mass distribution results in $\sim 4\times$ increase in the stacked X-ray luminosity from the hot CGM. Using empirical models to derive a permissible range of AGN and XRB luminosities in the MW-mass X-ray galaxy stack, we choose our forward model that best describes the hot CGM for the eROSITA observations. Our chosen model in the MW stellar mass bin is in good agreement with previous literature results. We find that at $\lesssim 40$ kpc from the galaxy center, the hot CGM from central galaxies and the X-ray point sources emission (from XRBs and AGNs) each account for $40-50\%$ of the total X-ray emission budget. Beyond $\sim 40$ kpc, we find that the hot CGM around satellites (probing their more massive host halos with mean $M_{\rm 200m}\sim 10^{14}\ \MSUN$) dominates the stacked X-ray surface brightness profile.
  }
  {The gas physics driving the shape of the observed hot CGM (in stellar-mass-selected X-ray stacking experiments) is tightly correlated with the underlying halo-mass distribution. This work provides a novel technique to constrain the AGN X-ray luminosity jointly with the radial hot CGM gas distribution within the halo using measurements from X-ray galaxy stacking experiments. Implementing this technique on other state-of-the-art simulations will provide a new ground for testing different galaxy formation models with observations.}
  
    \titlerunning{Modeling the hot CGM radial profiles in X-ray stacking experiments with eROSITA}
    \authorrunning{S. Shreeram}
   \maketitle

\section{Introduction}

The circumgalactic medium (CGM) plays a crucial role in a galaxy's evolution by directly tracing inflows and outflows of gas driven by various gravitational and nongravitational mechanisms (see \cite{faucher2023key} for an overview). The nongravitational mechanisms, such as stellar and active galactic nucleus (AGN) feedback, heat and cool the gas, impacting the evolution of star formation in the galaxy~\citep{donahue2022baryon}. The relative contributions from shock heating of the gas due to gravitational infall, stellar, and AGN heating are model-dependent in the current paradigm of galaxy formation and evolution. Particularly, the impact of stellar and AGN feedback depends on the host halo mass, where the former affects the halo masses $\lesssim 10^{12}\ \MSUN$, and the latter affects the $\gtrsim 10^{12}\ \MSUN$, respectively~\citep{Wechsler2018halo}. The pivotal point, where the relative contributions from stellar and AGN feedback are equally important, occurs at halo mass scales similar to the one of our Milky Way (MW), defining it as a crucial testing range for the models.

The interplay between feedback mechanisms and the galactic atmosphere results in a multiphase CGM, with the hot phase ($T \gtrsim 10^6$ K) typically being the most massive and volume-filling component (see review by \citealt{tumlinson2017circumgalactic}). The hot CGM radiates in the soft X-ray within the $0.2-2$ keV energy band for MW-mass halos due to thermal hot gas emission. There are various techniques to probe the hot CGM in X-rays via absorption~\citep{galeazzi2007xmm, bhattacharyya2023hot, mathur2023probing, Wijers2020eagle, bogdan2023circumgalactic} and emission~\citep{koutroumpa2007ovii, Bertone2010oviii, van2013owl, bogdan2013detection, bogdan2013hot, anderson2016deep, bogdan2017probing, li2017circum, das2019evidence, Zhang:2022un, ponti2023abundance, locatelli2024warm, Zheng2024ovii}. In particular, emission studies, combined with stacking, allow us to map the large-scale extent of the hot CGM to the halo's virial radius~\citep{anderson2015unifying, oppenheimer2020eagle, comparat2022erosita, Chadayammuri:2022us, zhang2024hot}. 

Given the advent of eROSITA~\citep{merloni2024srg}, there have been several studies exploiting the unprecedented statistics for stacking the X-ray emission at the position of optically selected galaxies, such as \cite{comparat2022erosita}, \cite{Chadayammuri:2022us}, and most recently \citet[hereafter Z24]{zhang2024hot}. \citetalias{zhang2024hot} represent the state-of-the-art hot CGM measurements for MW-mass galaxies, given the largest optical galaxy sample with the German half-sky eROSITA coverage in X-rays. They stack $415,627$ galaxies with photometric redshifts, Full$_{\rm phot}$, from the DESI Legacy Survey DR 9~\citep{dey2019overview, zou2019photometric, zou2022photometric} and $30,825$ central galaxies with spectroscopic redshifts from the SDSS DR7 Main Galaxy Sample~\citep{strauss2002spectroscopic, abazajian2009seventh}. The latter, the SDSS-based central galaxy sample, has the advantage of spectroscopic information, allowing for galaxies to be classified into centrals and satellites with halo mass information~\citep{tinker2021self}. Therefore, \citetalias{zhang2024hot} retrieve the X-ray surface brightness profile from the hot CGM by empirically modeling the impact due to satellites, AGNs, and XRB emissions. However, the former, the DESI Legacy survey-based galaxy sample, cannot be classified into centrals and satellites, given the limitations in photometric redshift, making the modeling of this dataset challenging. To exploit the highest signal-to-noise data (a factor of $13.5$ times more statistics than the spectroscopic sample) to date, in this work, we embark on constructing a forward model to disentangle the hot CGM radial profiles from the X-ray stack of optically selected galaxies.
  
Among the dominating sources of contamination in X-ray stacking experiments at MW-mass galaxies are \textit{(1)} the AGN and X-ray binary~(XRB) population of galaxies~\citep{Biffi:2018wj, Vladutescu-Zopp:2023vo}, and \textit{(2)} the effect of having satellite galaxies in the stacking sample, where these satellites contribute to the averaged X-ray stack with their more massive host galaxy within which they are embedded in the large-scale structure (see e.g.,~\citealt{shreeram2025quantifying, weng2024physical}). Given that we are modeling a photometric galaxy sample in this work, we do not have the classification of galaxies into centrals and satellites due to limitations in the (photometric) redshift accuracy for the galaxies in large optical surveys. Therefore, we use our TNG-based forward model fit for the magnitude of the satellite galaxy contribution by setting the normalization of the surface brightness profile free when fitting to data. The contaminating effect of the satellites was quantified in detail in \cite{shreeram2025quantifying}, where we find that this effect becomes increasingly significant in the stacking sample as the stellar mass decreases. When conducting blind X-ray stacking analysis at the positions of optically selected galaxies, where centrals and satellites are unclassified, the inclusion of satellites implies that the total measured X-ray surface brightness profile comprises (i) the intrinsic hot gas emission around truly central galaxies and (ii) the contamination of hot gas emission measured around satellites. We emphasize that the latter does not correspond to the emission intrinsic to the satellites as the more massive host (central) galaxy in the vicinity of the satellite dominates the emission, resulting in a negligible contribution of the intrinsic satellite emission (see e.g., \citealt{rohr2024sat}).

This paper presents a forward model for the stacked galaxy profile comprising the X-ray emitting gas and point source emission. 
We use the lightcone built with TNG300 in~\cite{shreeram2025quantifying} to construct mock galaxy catalogs representing the observations. From our TNG-based mock galaxy catalog, we predict the hot gas CGM profile contribution to the X-ray galaxy stack from central and satellite galaxies. We parameterize the normalization of the hot gas contribution from satellites, $\mathcal{N}_{\rm sat}$. The hot gas prediction for central galaxies from our forward model with TNG is left unchanged. As for the point-like source contributions from AGNs and XRBs, we do not use the simulations, which are quite uncertain in their predicted instantaneous AGN activity, but instead we describe their contribution with a normalization parameter, $\mathcal{N}_{\rm ps}$, and fix the radial shape using the point spread function (PSF) of eROSITA. This leaves us with two fitting parameters in our model: the contribution of point sources in the stack, $\mathcal{N}_{\rm ps}$, and the normalization of the satellite X-ray surface brightness profile, $\mathcal{N}_{\rm sat}$. We fit our forward model to the mean X-ray surface brightness profile reported \citetalias{zhang2024hot}, obtained by stacking MW-mass galaxies. We obtain the point source luminosity from our fitting analysis, and we interpret our results by comparing them with independent empirical models of AGN luminosity functions. 

The paper is organized as follows. Sec.~\ref{sec:dataZ24} expands on the observational data used for modeling in this study. Sec.~\ref{sec:fmodel} describes the forward model built in this work; details on the point source and hot gas component are provided in Sec.~\ref{subsec:fmodel_point_src} and Sec.~\ref{subsec:fmodel_hot}, respectively, and details on the generation of mock galaxy catalogs in Sec.~\ref{subsec:mock}. Sec.~\ref{sec:results} interprets and discusses the results, and Sec.~\ref{sec:summary} summarizes the main findings of this work.

\section{Data}
\label{sec:dataZ24}

This study uses the results from~\citetalias{zhang2024hot}, where they use optical data from the Legacy Survey Data release 9~(\citealt{dey2019overview}), and X-ray data from the first four \textit{SRG}/eROSITA All-Sky Surveys (eRASS:4) within the western galactic hemisphere~\citep{merloni2024srg}. Here, we summarize the relevant details from~\citetalias{zhang2024hot} to motivate the forward-model setup, as is described in Sec.~\ref{sec:fmodel}.

The LS DR9 overlaps by $9,340$ deg$^2$ with the western Galactic hemisphere. \citetalias{zhang2024hot} define the Full$_{\rm phot}$ sample containing $1,677,909$ galaxies, which is based on the LS DR9 galaxy catalog from \cite{zou2019photometric, zou2022photometric}, where they provide the galaxy properties. The stellar masses, $M_\star$, with uncertainties of $\sim 0.2$ dex are provided in the range $9.5 <\log M_\star < 11.5$ and the photometric redshifts, $z_{\rm phot}$, with uncertainty $\Delta z_{\rm phot} \lesssim 0.03$, spans the range $0.01< z_{\rm phot} < 0.4$ (see Tab. 3 in \citetalias{zhang2024hot} for further classification into stellar mass bins). We focus here on the modeling of the X-ray emission from MW-like galaxies, defined by the stellar mass bin $10.5 < \log_{10} M_\star < 11.0$, and located in the redshift range $0.02 < z_{\rm phot} < 0.17$; containing $415,627$ galaxies and median stellar mass and redshift of $5.5\times 10^{10}\ \MSUN$ and $z_{\rm phot} = 0.14$, respectively.

The details on the data analysis pipeline to obtain the observed X-ray surface brightness profiles, based on the LS DR9 (Full$_{\rm phot}$) galaxies, are provided in~\citetalias{zhang2024hot}. They stack the X-ray data in different stellar mass bins following the method from \cite{comparat2022erosita}. The first step is to generate X-ray event cubes around every galaxy in the sample within $3$ Mpc. The events within the cube are assigned a physical radial distance from the source, $R_{\rm kpc}$, along with the exposure time,~$t_{\rm exp}$, effective collecting area, $A_{\rm eff}$, and the rest frame energy of the event around the source,~$E_{\rm rest}$. The X-ray surface brightness profile around the galaxy is obtained, with additional correction factors due to absorption and area loss if masking of sources is applied (see Eq.~1 in \citetalias{zhang2024hot}). We use the background-subtracted profiles for modeling in this work (for details on the background treatment in the data, see Appendix A in \citetalias{zhang2024hot}).

\section{Forward model built in this work}
\label{sec:fmodel}

This section explains the forward model we built to fit the observed X-ray surface brightness profile reported in \citetalias{zhang2024hot}. The mean X-ray surface brightness profile obtained from X-ray stacking of galaxies on their optically detected positions is expressed as
\begin{equation}\label{eq:full_Sx}
    {\mathbf{S}}_{\rm X,\ total}(r) = {\mathbf{S}}_{\rm X,\ hot\ gas}(r)   +  {\mathbf{S}}_{\rm X,\ point-source}(r) ,  
\end{equation}
where ${\mathbf{S}}_{\rm X,\ hot\ gas}[r]$ is the X-ray contribution to $\mathbf{S}_{\rm X,\ total}(r)$ from hot gas (further detailed in Sect.~\ref{subsec:fmodel_hot}) and ${\mathbf{S}}_{\rm X,\ point-source}[r]$ is the X-ray contribution from AGNs and XRBs (Sect.~\ref{subsec:fmodel_point_src}). We then introduce the forward models constructed in this work in Sec.~\ref{subsec:mock}.

\subsection{The point source component}
\label{subsec:fmodel_point_src}

The ${\mathbf{S}}_{\rm X,\ point-source}[r]$ component is expressed as
\begin{equation}\label{eq:Lx_ps}
    {\mathbf{S}}_{\rm X,\ point-source}(r) = \mathcal{N}_{\rm ps} \  {\rm PSF}(z, r)  
,\end{equation}
where $\mathcal{N}_{\rm ps}$ is the normalization of the point-source component and ${\rm PSF}[z, r] $ is the shape of the point-source component as defined by the survey-averaged eROSITA PSF. Here, the shape of the mean PSF was obtained by converting from angular to physical scale [kiloparsecs] using the redshifts of the galaxies in the stacking sample. We constrained $\mathcal{N}_{\rm ps}$ by fitting the observations.

We did not use the TNG300 outputs to predict the contributions from AGNs and XRBs for the following reasons. As is presented in \cite{Habouzit2019xlf}, the X-ray luminosity function (XLF) for AGNs in TNG shows an overproduction of faint AGNs at $z=0$, a common problem in hydrodynamical cosmological simulations due to poorly resolved sub-grid feedback prescriptions \citep{sijacki2015illustris, volonteri2016cosmic, rosas2016supermassive, Biffi:2018wj}. Additionally, in TNG, the bright end underpredicts the XLF compared with observations due to the over-efficient kinetic mode of feedback prescriptions with TNG~\citep{Habouzit2019xlf}. Given these discrepancies between the hydrodynamical simulations and observations, using the TNG-based predictions for the X-ray AGN contribution would be unreliable for the purpose of this work. As for XRBs, whose prediction depends on the star formation rates (SFRs) of galaxies (more details in Sec.~\ref{subsec:XRB}), using the TNG SFR values for the mock galaxies to represent observations requires extreme care, given how sensitive the SFR is to the definition of quenched galaxies, the physical apertures used for measurement of the SFR, and the mass resolution of the simulation~\citep{donnari2019star}. Additionally, the quenched fraction of satellite galaxies is overestimated in the stellar mass range of $M_\star \in 10^{10-11} \ \MSUN$ at $z=0$~\citep{donnari2021quenched}, which entails the mass range of interest in this work. These caveats impede us from reliably predicting meaningful estimates for point source contributions intrinsic to the TNG model that can be directly compared with observations.

In this work, we use independent empirical estimates (Sec.~\ref{subsec:XRB} and Sec.~\ref{subsec:AGN}) to predict the allowed range of the mean X-ray luminosity from XRBs and AGNs, respectively. Since the estimates from these empirical models are independent of TNG, we use them to inform our forward models for the permitted values of point source luminosities, as is shown in Fig.~\ref{fig:exclusion_models}.

\begin{figure*}

    \centering
    \begin{minipage}{0.5\textwidth}
        \includegraphics[width=\linewidth]{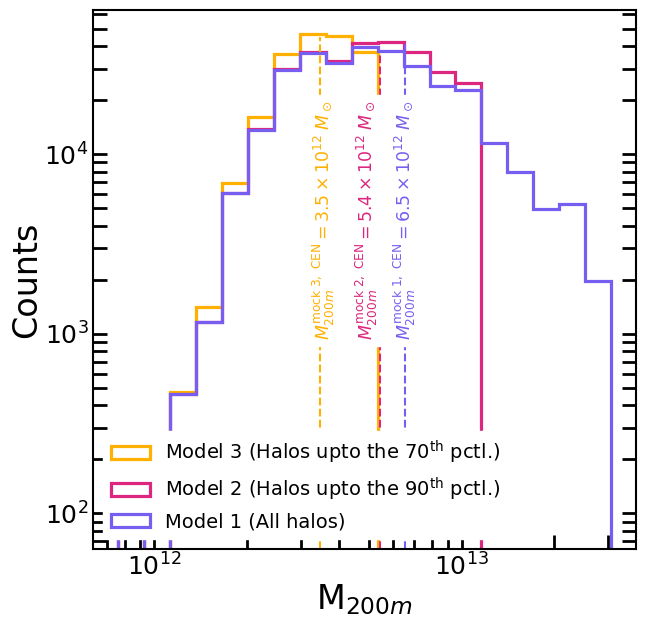}
    \end{minipage}%
    \centering
    \begin{minipage}{0.5\textwidth}
        \includegraphics[width=\linewidth]{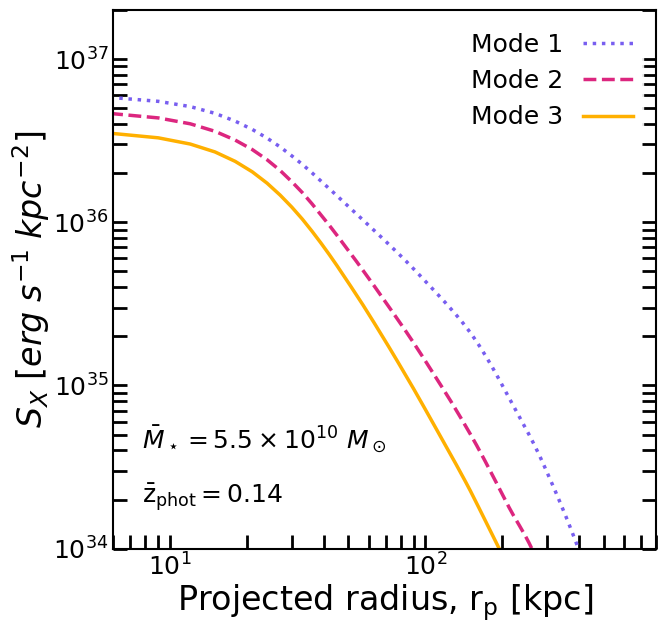}
    \end{minipage}%
    \caption{Forward models constructed in this work for the hot CGM from central galaxies by varying the underlying halo mass distribution. {\it Left panel}: Purple halo mass distribution (Model 1) obtained from the mock central galaxy catalog -- constructed with the TNG300 lightcone (LC-TNG300) from \cite{shreeram2025quantifying} -- for the X-ray stack from \cite{zhang2024hot} that uses optically detected galaxies with photometric redshifts (Full$_{\rm phot}$) from LS DR9~\citep{dey2019overview}. Note that the mock galaxy catalog is generated by matching LC-TNG300 to Full$_{\rm phot}$ in stellar mass and redshift (see details in Sec.~\ref{subsec:mock}); the median stellar mass and redshift of the Full$_{\rm phot}$ (and our mock catalogs) are $5.5\times 10^{10} \ \MSUN$ and $0.14$, respectively. The underlying halo mass distribution of the Full$_{\rm phot}$ optical dataset is unknown. The pink distribution (Model 2; with mean $M_{\rm 200m} = 5.4\times 10^{12}\ \MSUN$) discards the top $10\%$ most massive halos before the generation of the mock galaxy catalog. The yellow distribution (Model 3; with mean $M_{\rm 200m} = 3.5\times 10^{12}\ \MSUN$) discards the top $30\%$ most massive halos before the generation of the mock galaxy catalog. {\it Right panel}: Corresponding X-ray surface brightness profiles in the $0.5-2$ kev energy band (for details on their generation see Sec.~\ref{subsec:fmodel_hot}) for the three mock galaxy catalogs with different halo mass distributions, which are shown in the left panel. The profiles are convolved with the eROSITA PSF, and they represent the Full$_{\rm phot}$ dataset in the stellar mass and redshift plane. Nevertheless, due to the impact of the underlying halo mass distribution, the shape and normalization of the hot CGM profiles are impacted, where discarding the most massive halos from the underlying halo distribution results in steeper profiles with lower normalizations.}
    \label{fig:halo-models}
\end{figure*}

\subsection{The hot gas component}
\label{subsec:fmodel_hot}

In this work, we modeled the hot gas emission using the TNG300 hydrodynamical simulations \citep{pillepich2018first, marinacci2018first, naiman2018first, nelson2015illustris, springel2018first}; we used TNG300 to construct a lightcone and generate mock X-ray observations, as presented in~\cite{shreeram2025quantifying}. Here, we summarize the most important features. We used the IllustrisTNG cosmological hydrodynamical simulation with the box of side length $302.6$ Mpc~\citep[TNG300;]{nelson2019illustristng}\footnote{\url{ http://www.tng-project.org} }; this box size allows us to map the hot CGM around MW-mass analogs embedded in the large-scale structure. TNG300 contains $2500^3$ dark matter particles, with a baryonic mass resolution of $1.1\times 10^7 \rm \ M_{\odot}$ (resulting in $\gtrsim 10^3$ particles at MW-mass galaxies), a comoving value of the adaptive gas gravitational softening length for gas cells of $370$ comoving parsec  (allowing us to resolve the X-ray gas from $\sim 5$ kpc from the halo center), gravitational softening of the collisionless component of $1.48$ kpc, and dark matter mass resolution of $5.9\times 10^7\ \rm M_{\odot}$. The TNG simulations adopt the \citetalias{ade2016planck} cosmological parameters. The TNG300 lightcone, LC-TNG300, was constructed with the box remap technique~\citep{carlson2010embedding}, {using the \texttt{LightGen} code\footnote{The code to generate lightcones from TNG is publicly available at \url{https://github.com/SoumyaShreeram/LightGen/}}}. LC-TNG300 spans across redshifts $0.03 \lesssim z \lesssim 0.3$; this range is motivated by observations (e.g., \citealt{comparat2022erosita, Chadayammuri:2022us, zhang2024hot}). It goes out to $1231\rm \ cMpc$ along the x axis, subtending an area of $47.28\rm \ deg^2$ on the sky in the y-z plane. The physical properties of the distinct halos and subhalos within the TNG300 lightcone are obtained by the \textsc{subfind} algorithm~\citep{springel2001populating, dolag2009substructures}. \textsc{subfind} detects gravitationally bound substructures, equivalent to galaxies in observations, and also provides us with a classification of subhalos into centrals and satellites, where centrals are the most massive substructure within a distinct halo. For the MW-mass bin\footnote{This paper defines the stellar mass used from TNG300 as the mass within twice the stellar half-mass radius.}, $\rm M_\star = 10^{10.5-11} \MSUN$, we have $5,109$ centrals and $2,719$ satellites, resulting in a total simulated galaxy catalog with $7,828$ galaxies (subhalos).

The X-ray photons were simulated within the LC-TNG300 in the $0.5-2.0$ keV intrinsic band with \texttt{pyXsim}~\citep{zuhone2016pyxsim}, which is based on \textsc{phox}~\citep{Biffi:2013uh, biffi2018origin}, by assuming an input emission model where the hot X-ray emitting gas is in collisional ionization equilibrium. The spectral model computations of hot plasma use the Astrophysical Plasma Emission Code, \textsc{apec}\footnote{APEC link \url{https://heasarc.gsfc.nasa.gov/xanadu/xspec/manual/node134.html}} code~\citep{smith2001collisional} with atomic data from \textsc{atomdb} v3.0.9~\citep{foster2012updated}. This model uses the plasma temperature of the gas cells (in keV),  the redshift $z$, and metallicity; \cite{shreeram2025quantifying} assume a constant metallicity of $0.3\ Z_\odot$ for the generation of X-ray events. The X-ray events use the solar abundance values from~\cite{anders1989abundances}. The events are generated by assuming a telescope with an energy-independent collecting area of $1000$ cm$^2$ and an exposure time of $1000$ ks. The photon list is generated in the observed frame of the X-ray emitting gas cells and is corrected to rest frame energies. Finally, the photons generated by the gas cells are projected onto the sky.

We obtained X-ray radial surface brightness profiles in the $0.5-2.0$ keV band for all galaxies. Given that \textsc{subfind} provides us with an accurate classification of galaxies into centrals and satellites, we distinguish the hot gas component into X-ray emissions around centrals and satellites. For central galaxies, the profiles represent the hot gas emission around them; however, for satellite galaxies, the profiles probe the hot gas emission of the more massive host halo in the vicinity. 

We convolved the individual X-ray surface brightness profiles from LC-TNG300-based mock galaxies catalogs with the eROSITA PSF \citep{merloni2024srg}. The PSF convolved mean X-ray surface brightness profile from hot gas, ${\mathbf{S}}_{\rm X,\ hot\ gas}[r]$, is expressed as follows:
\begin{multline}\label{eq:hot_gas}
\centering
    \mathbf{S}_{\rm X,\ hot\ gas}(r) = f_{\rm cen}\ {S}_{\rm X,\ cen} + 
    \mathcal{N}_{\rm sat} 
    \times f_{\rm sat}\ {S}_{\rm X,\ sat}
    ,
\end{multline}
where $ {S}_{\rm X,\ cen}$ is the TNG-based prediction for the hot gas around central galaxies and ${S}_{\rm X,\ sat}$ corresponds to the hot gas around satellites. After matching LC-TNG300 with Full$_{\rm phot}$ in stellar mass and redshift, the mock galaxy catalogs fix the fraction of centrals, $f_{\rm cen}$, and satellites, $f_{\rm sat}$. $\mathcal{N}_{\rm sat} $ is the factor by which the mock prediction from ${S}_{\rm X,\ sat}$ is rescaled to match the observations, thereby renormalizes the ${S}_{\rm X,\ sat}$ of the TNG300-based prediction; $\mathcal{N}_{\rm sat}$ is the only free parameter in the hot gas emission component.

The motivation behind introducing the renormalization parameter, $\mathcal{N}_{\rm sat}$, for fitting the forward model prediction for  ${S}_{\rm X,\ sat}$ with observations is as follows. The TNG-based prediction for ${S}_{\rm X,\ sat}$ from the mock catalogs (for a given $f_{\rm sat}$) is $\sim 5-7\times$ brighter than the Full$_{\rm phot}$ stack. \cite{shreeram2025quantifying} find that the shape of the X-ray radial surface brightness profile from satellite galaxies (hosted by massive halos) is unaffected by $f_{\rm sat}$ in the galaxy sample. This is because the halo masses making up the average profile from satellite galaxies, whose $M_\star \in 10^{10.5-11} \ \MSUN$, are dominated by host (central) halos with mean $M_{\rm 200m}\sim 10^{14}\ \MSUN$. Therefore, by changing the normalization of the ${S}_{\rm X,\ sat}$, we effectively damp the normalization of the X-ray thermal gas contribution from the most massive clusters in the simulation. This is justified given that the hot gas fraction from TNG is overpredicted at halo masses above $M_{\rm 500c} \gtrsim 10^{13.5}\ \MSUN$, as shown in Fig.~6 in \cite{popesso2024hot}. This is also reflected in the $L_{\rm X}-M_{\rm 500c}$ relation shown in \cite{zhang2024hot2} and \cite{popesso2024average}.

We emphasize that the X-ray surface brightness profile prediction from central galaxies, $ {S}_{\rm X,\ cen}$, which represents the CGM physics of interest in this work, is untouched. We predict multiple CGM profiles by changing the host halo mass distribution of the central galaxies and propagating it through our pipeline to generate mock galaxy catalogs for each halo distribution considered, as detailed in the following section. Note that the stellar mass and redshift distributions are the same for all three models.

We restricted our analysis to the MW-mass stellar bin due to both observational and modeling constraints. While comparisons across a broader stellar mass range are desirable, for lower-mass galaxies, there is currently no suitable X-ray stacking data available. For higher-mass galaxies, the existing stacked X-ray measurements (e.g., from \citetalias{zhang2024hot}) extend to redshifts beyond the coverage of our lightcone, LC-TNG300, which is built from the TNG300 simulation. Extending the analysis to these massive galaxies would require constructing a new lightcone from a larger-volume simulation (e.g., FLAMINGO, MTNG), generating matched mock X-ray observations, and repeating the full forward-modeling pipeline introduced here, an effort that warrants a future study. In contrast, the MW-mass bin provides both observational data within the redshift range of LC-TNG300, making it the optimal case for robust model–data comparison and the hot CGM signal retrieval.

\begin{table*}[]
\centering
\caption{Summary of the best-fit parameters (see Eq.~\ref{eq:full_Sx}-\ref{eq:hot_gas}) and derived quantities (luminosity values) obtained from fitting the three forward models from this work to the Full$_{\rm phot}$ X-ray surface brightness profile.\label{tab:best-fit-params0}}
\resizebox{0.8\textwidth}{!}{%
\begin{tabular}{lllllll}
\multicolumn{1}{c}{} & \multicolumn{1}{c}{\begin{tabular}[c]{@{}c@{}}Best-fit $\mathcal{N}_{\rm sat}$\\ {[}10$^{34} \ \frac{\rm ergs}{\rm s\ kpc^2}${]}\end{tabular}} & \multicolumn{1}{c}{$f_{\rm sat}$} & \multicolumn{1}{c}{\begin{tabular}[c]{@{}c@{}}Best-fit $\mathcal{N}_{\rm ps}$\\ $\frac{\rm ergs}{\rm s\ kpc^2}$\end{tabular}} & \multicolumn{1}{c}{\begin{tabular}[c]{@{}c@{}}$L_{X,\rm \ PS}$\\ {[}ergs/s{]}\end{tabular}} & \multicolumn{1}{c}{\begin{tabular}[c]{@{}c@{}}$L_{X,\rm \ cenCGM}$\\ {[}ergs/s{]}\end{tabular}} & \multicolumn{1}{c}{$\chi^2_{\rm red}$} \\ \hline
Model 1 & $ 2.75_{-0.12}^{+0.12}$ & 0.31 & $ 7.6_{-1.6}^{+2.2}\times 10^{36}$ & $ 7.851_{-0.274}^{+0.028}\times 10^{39}$ & $ 6.58_{-0.94}^{+0.27}\times 10^{40}$ & $ 0.44$ \\
Model 2 & $ 2.95_{-0.13}^{+0.14} $ & 0.33 & $ 1.1_{-0.2}^{+0.2}\times 10^{37}$ & $ 1.195_{-0.042}^{+0.003}\times 10^{40}$ & $ 2.76_{-0.19}^{+0.06}\times 10^{40}$ & $ 1.32$ \\
Model 3 & $ 1.51_{-0.073}^{+0.07} $ & 0.56 & $ 1.5_{-0.2}^{+0.2}\times 10^{37}$ & $ 1.576_{-0.063}^{+0.001}\times 10^{40}$ & $ 1.69_{-0.91}^{+0.28}\times 10^{40}$ & $1.89$
\end{tabular}%
}
\tablefoot{For every model, we present the best-fit $\mathcal{N}_{\rm sat}$: renormalization of the ${S}_{\rm X,\ sat}$ (of the TNG300-based prediction), $f_{\rm sat}$: the fraction of satellites in the mock galaxy catalog (see descriptions of the mock catalogs in Sec.~\ref{subsec:mock}), $\mathcal{N}_{\rm ps}$: the normalization of the point-source component (see Eq.~\ref{eq:Lx_ps}), $L_{X,\rm \ PS}$: the X-ray luminosity obtained by integrating the point source component, $L_{X,\rm \ cenCGM}$: the X-ray luminosity obtained by integrating the central galaxies hot CGM component, $\chi^2_{\rm red}$: the reduced $\chi^2$ statistic for the model, obtained by using $21-2$ degrees of freedom.}
\end{table*}

\subsection{Mock galaxy catalogs}
\label{subsec:mock}

We next used the LC-TNG300 galaxy catalog to construct a mock galaxy sample for the LS DR9 Full$_{\rm phot}$ galaxies. We matched every one of the $415,627$ galaxies in the Full$_{\rm phot}$ sample with a galaxy from LC-TNG300 in redshift and stellar mass. By construction, the simulated LC-TNG300 galaxies follow the same stellar mass and redshift distribution as the observational sample. The mock sample predicts the mean X-ray surface brightness profile for gas emitted around centrals and satellites.

\begin{figure}

    \includegraphics[width=\linewidth]{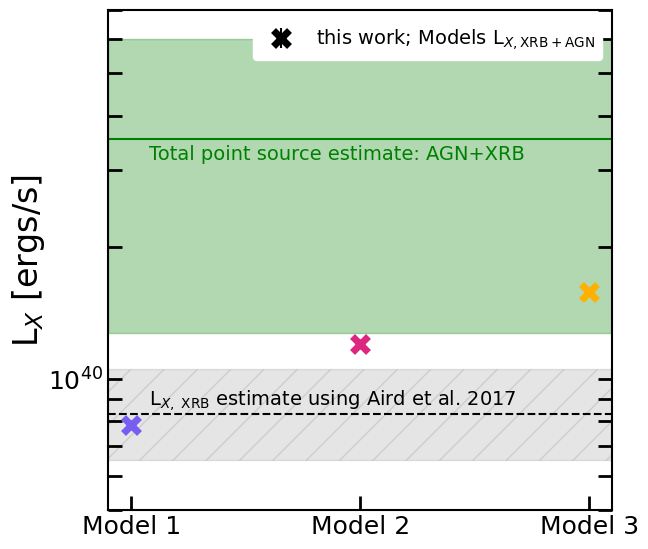}
    \caption{Comparison of the mean point source (AGN and XRB) luminosities from our three forward models (crosses, based on the different halo distributions shown in Fig.~\ref{fig:halo-models}) with the empirically allowed range of XRB and total point source luminosities, as shown by the hatched gray region and the shaded green region, respectively. We estimate the contribution due to XRB emission using the \cite{aird2017x} model. For estimating the AGN luminosity budget, ${L}_{X,\ \rm AGN}$, we used the \cite{aird2013primus} model for the incidence rate distribution as a function of the $L_{X, \rm \ AGN}^{2-10\ \rm keV}$ keV. To convert the $2-10$ keV luminosity distribution in the $0.5-2$ keV band, we used an empirical obscuration model from \cite{Comparat2019agnmodel}. For more details, see the text of Sec.~\ref{subsec:XRB} and~\ref{subsec:AGN}. This comparison favors model~$3$, shown by the yellow cross, where the hot CGM component allows for a point source component with luminosity that agrees with empirical estimates from the low-redshift Universe.}
    \label{fig:exclusion_models}
\end{figure}

In this work, we also tested the impact of the underlying halo mass distribution on the CGM physics. Therefore, we additionally generated two other mock galaxy catalogs using LC-TNG300, matched in stellar mass and redshift; however, with different underlying halo mass distributions (see left panel of Fig.~\ref{fig:halo-models}). Consequently, we also emulated the corresponding X-ray surface brightness profiles by varying the halo distributions (see right panel of Fig.~\ref{fig:halo-models}). The differences between the three forward models are as follows:
\begin{itemize}
    \item Model $1$ leaves LC-TNG300 halo mass distribution unchanged (purple line in Fig.~\ref{fig:exclusion_models}), resulting in the mean halo mass of $M_{200m} = 6.5\times 10^{12} \MSUN$.
    \item Model $2$ changes the underlying central galaxy halo distribution  by excluding the $10\%$ most massive (central) halos from the original LC-TNG300 halo mass distribution (pink line in Fig.~\ref{fig:halo-models}). This results in the mean halo mass of $M_{200m} = 5.4\times 10^{12} \MSUN$.
    \item Model $3$ changes the underlying halo distribution by excluding the $30\%$ most massive (central) halos from the original LC-TNG300 halo mass distribution (yellow line in Fig.~\ref{fig:halo-models}). This results in the mean halo mass of $M_{200m} = 3.5\times 10^{12} \MSUN$.
\end{itemize}

By changing the underlying halo mass distributions for the fixed stellar mass bins, we are effectively altering the mean halo mass of our mock galaxy catalog. We note that for model~$3$, the resulting mean halo mass lies within the range predicted by various observational and empirical works that constrain the stellar-to-halo-mass relation (SHMR) at low redshift. For the mean stellar mass of our three mock samples ($M_{\star} = 5.5\times 10^{10}\ \MSUN$), studies typically find $M_{\rm halo} \sim 5\times 10^{11}\ \MSUN$~\citep{Taylor2020z_0_18} to $1$--$3\times 10^{12}\ \MSUN$~\citep{Leauthaud2012z_0_2, Coupon2015_z_0_8, Girelli2020shmr, behroozi2019universemachine}. {Importantly, different simulations predict different SHMRs~(see e.g.,~\citealt{wright2024baryon}), and this variation can introduce systematic biases when comparing predicted CGM emission at fixed stellar mass. If the simulated halo mass distribution does not accurately reflect that of the observational sample, the predicted X-ray signal may be systematically over- or under-estimated. For instance, using a simulation that associates higher halo masses at fixed stellar mass than the true sample would yield artificially elevated CGM emission predictions, potentially leading to an erroneous conclusion about a simulation–observation mismatch. This underscores the motivation for our approach: rather than adopting a single SMHM relation, we forward-modeled the halo mass distribution to make it consistent with the data, using observational constraints (further discussed in Sec.~\ref{subsec:XRB}-\ref{subsec:model3}).} Our framework thus enables posterior constraints on the halo mass distribution associated with the Full$_{\rm phot}$ dataset, based on the TNG-informed models developed in this work. Given the large scatter in halo mass at fixed stellar mass \citep[e.g.,][]{moster2020emerge}, and the strong dependence of CGM surface brightness profiles on this distribution (see right panel of Fig.~\ref{fig:halo-models}), a rigorous treatment of halo demographics is essential for robust comparisons between simulations and stacked X-ray observations.

\section{Results and discussion}
\label{sec:results}

We fit the data from \citetalias{zhang2024hot} with our three forward models, which contain the hot gas component and the point source component, as shown in Eq.~\ref{eq:full_Sx}-\ref{eq:hot_gas}. The three models emulate different X-ray surface brightness profiles for different halo mass distributions (Sec.~\ref{subsec:mock} and Fig.~\ref{fig:halo-models}). We implemented the Markov chain Monte Carlo~\citep[MCMC]{hastings1970monte} method to determine the posterior probability distributions of the two free parameters of our models: $\mathcal{N}_{\rm sat}$ and $\mathcal{N}_{\rm ps}$. The parameters were obtained using the Affine-Invariant Ensemble Sampler algorithm in \texttt{emcee}~\citep{foreman2013emcee}. We assumed a Gaussian likelihood function and uniform priors on $\mathcal{N}_{\rm sat}\in (0.005, 1000)\times 10^{35}$, and $\mathcal{N}_{\rm ps} \in (0.5,\ 550)\times 10^{35} {\rm erg/s/kpc^2}$. For the three forward models constructed in this work, we show the most likely values of the free parameters in Tab.~\ref{tab:best-fit-params0}. We computed the luminosities from the hot gas around centrals, satellites, and point sources within $R_{\rm 500c}$\footnote{$R_{\rm 500c}$ is the radius at which the density of the halo is  $500\times$ the critical density of the Universe.}. Fig.~\ref{fig:exclusion_models} shows the mean point source luminosities we obtain for the three models implemented in this work (purple, pink, and yellow crosses). We compare our results with independent predictions on the expected luminosity from XRBs and AGNs around MW-mass galaxies using current empirical models in the literature. Sec.~\ref{subsec:XRB} and Sec.~\ref{subsec:AGN} describe how we obtain these estimates shown in Fig.~\ref{fig:exclusion_models} for expected luminosity from XRBs and AGNs around MW-mass galaxies.

\subsection{Predicting the X-ray emission from XRBs}
\label{subsec:XRB}

The XRB emission, which is the X-ray emission from the binary component of stellar populations in normal galaxies, is divided into high-mass X-ray binaries (HXRBs) and low-mass X-ray binaries (LXRBs); see review by \cite{fabbiano2006populations}. The average XRB emission from a normal galaxy is characterized by scaling laws, where the former HXRB population scales with the recent star formation rate (SFR) in the galaxy~\citep{Grimm2003hmxb, Shtykovskiy2005smc,mineo2012x}. In contrast, LXRB emission spans longer timescales, tracing the stellar mass of the galaxy~\citep{gilfanov2004low, boroson2011revisiting, Zhang2012lxrb, lehmer2019x}. The total XRB emission from extragalactic objects is distributed on the scale of the stellar body; however, for an instrument with a $30$ arcsec PSF like eROSITA, it is unresolved and appears as a point source.
\cite{aird2017x} and \cite{lehmer2016evolution} provide simple empirical recipes by parameterizing the total X-ray luminosity from XRBs as a function of both the SFR and stellar mass, M$_\star$, of the galaxy,
\begin{equation}\label{eq:Lx_XRB}
    L_{X,\ \rm XRB} = L_{X,\ \rm LXRB} + L_{X,\ \rm HXRB} = \alpha (1 + z)^\gamma M_\star + \beta (1+z)^\delta {\rm SFR}^\theta,
\end{equation}
where $\alpha$, $\beta$, $\gamma$, $\delta$, and $\theta$ are fitting constants. \cite{aird2017x} report the following best-fitting values: $\log \alpha = 28.81 \pm 0.08$, $\gamma = 3.90 \pm 0.36$, $\log \beta = 39.50 \pm 0.06$, $\delta = 0.67 \pm 0.31$ and $\theta = 0.86 \pm 0.05$.

We quantified the contribution of the total $L_{X,\ \rm XRB}$ in the LS DR9 Full$_{\rm phot}$ galaxy catalog using the model from \cite{aird2017x}. Since we would later use these estimates to inform our forward models for the allowed range of point source luminosities, we adopted a TNG-independent method of predicting $L_{X,\ \rm XRB}$ unbiasedly (other reasons for not using TNG are also detailed in Sec.~\ref{subsec:fmodel_point_src}). We used \textsc{uchuu}, a suite of ultra-large cosmological N-body simulations~\citep{ishiyama2021uchuu}, with the galaxy catalog from \textsc{universemachine}~\citep{behroozi2019universemachine} to construct a mock for the Full$_{\rm phot}$ galaxy sample. The SFRs from \textsc{universemachine} were calibrated to reproduce observations. We used the half-sky lightcone, constructed in the procedure detailed in \cite{ComparatEckertFinoguenov_2020}, to build the mock galaxy catalog. The mocks were generated similarly to \citetalias{zhang2024hot}, ensuring that the galaxy stellar mass function of the LS DR9 Full$_{\rm phot}$ galaxy catalog was reproduced. Therefore, they could reliably be used for the purpose of this study. We applied Eq.~\ref{eq:Lx_XRB} on the mocks to estimate the contribution of XRBs in the galaxy stack, given the stellar masses and SFR of the mock galaxies. With these ingredients, we predict that the contribution from XRBs alone to be $L_{X, \rm \ XRB} = (8.2_{-1.7}^{+2.2})\times 10^{39}$~ergs/s, represented by the hatched gray region in Fig.~\ref{fig:exclusion_models}\footnote{We note that the X-ray scaling relation from \citet{lehmer2019x} predicts $L_{X,\rm XRB} = (5.1_{-0.9}^{+1.6})\times 10^{39}$~erg/s for the Full$_{\rm phot}$ galaxy catalog. Nevertheless, the predictions from using the \cite{aird2017x} and \cite{lehmer2019x} scaling relations are consistent with each other, and we have verified that adopting the Lehmer relation would not change our results.
}.

\begin{figure}
    \includegraphics[width=\linewidth]{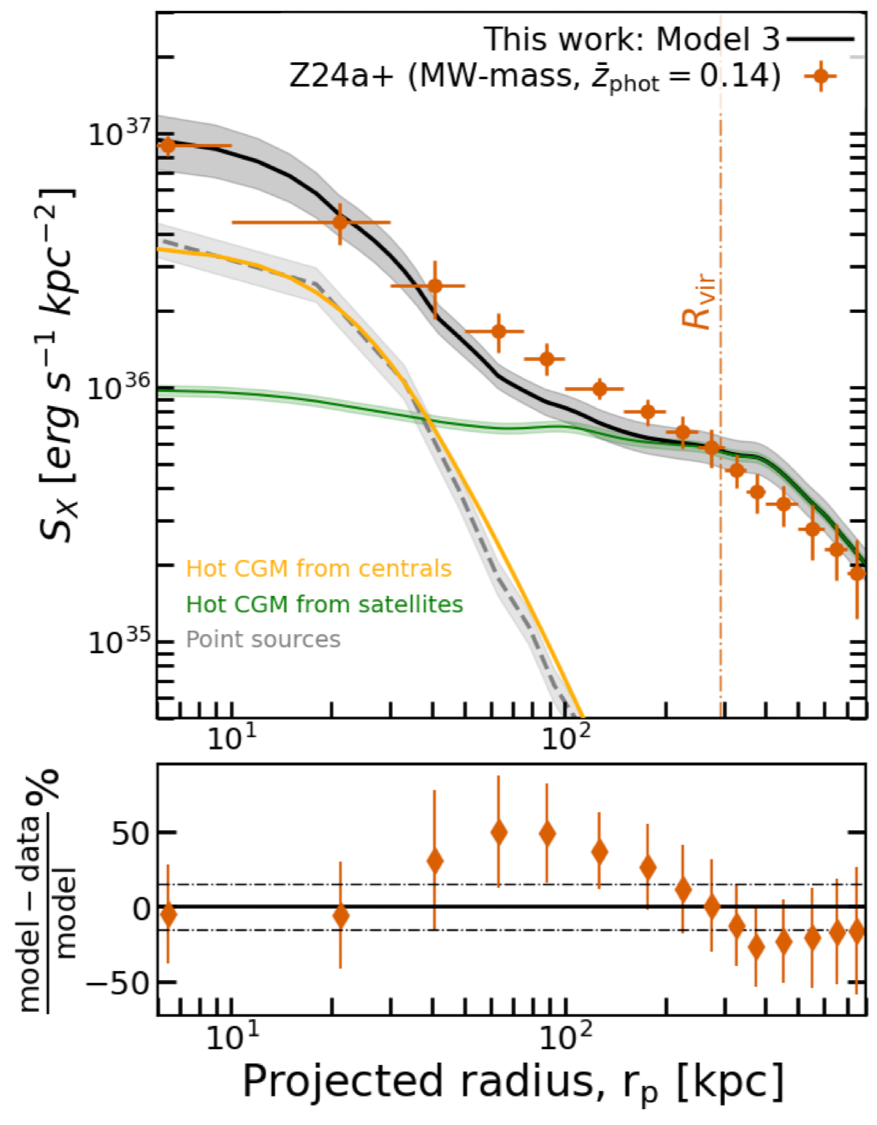}
    \caption{Decomposition of the X-ray stack of the galaxies in the photometric sample, Full$_{\rm phot}$,  into contributions from hot gas events (centrals and satellites hosted by more massive host halos) and point sources (AGNs and XRBs). The orange data points from \citetalias{zhang2024hot} are described with the model from this work (shown by the solid black line). The dash-dotted orange line at $292$ kpc corresponds to the virial radius of the observational sample. The model is composed of the following: the hot CGM from central galaxies (yellow), the events around satellites probing the hot gas of their more massive host halos (green), and X-ray events from unresolved and resolved point-like sources comprising AGNs and XRBs (gray). The bottom panel shows the percentage deviation of the best-fit forward model from the data. The dash-dotted lines show the $15\%$ level.}
    \label{fig:best-fit}
\end{figure}

\begin{figure}

    \includegraphics[width=\linewidth]{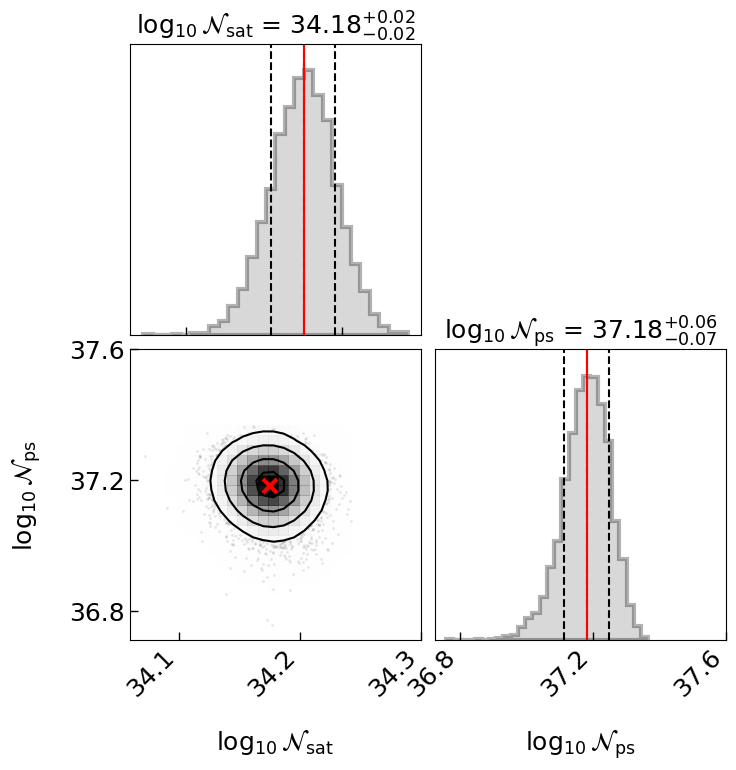}
    \caption{Posterior probability distributions of the renormalization factor of the ${S}_{\rm X,\ sat}$ profile: $\mathcal{N}_{\rm sat}$, and the normalization of the point source component: $\mathcal{N}_{\rm ps}$, which are obtained by fitting the forward-model $3$ from this work to the Full$_{\rm phot}$ data points from \citetalias{zhang2024hot} shown in Fig.~\ref{fig:best-fit}. The vertical red lines in the diagonal plots correspond to the most likely value; the respective values are mentioned in the titles (refer to Tab.~\ref{tab:best-fit-params0}). The dashed black lines are the $68\%$ confidence interval of the marginalized distribution of the free parameters. The contour plot marks the most likely values with the red cross, and the contours correspond to the $68\%, 95\%$, and $99.7\%$ confidence intervals.}
    \label{fig:best-fit-posterior}
\end{figure}

\subsection{ Predicting expected ${L}_{X,\ \rm AGN} $ for MW-mass galaxies using an empirical model for the low-redshift Universe}
\label{subsec:AGN}

X-ray emission from AGNs originates around an accreting supermassive black hole (see \citealt{brandt2015cosmic} for a review), appearing as a point-like X-ray source with eROSITA. We proceed to use the empirical model from \citet[hereafter A13]{aird2013primus} to estimate the $L_{X,\rm \ AGN}$ for a mean stellar mass $M_\star = 10^{10.7} \ \MSUN$. \citetalias{aird2013primus} provide a model for the probability of a galaxy hosting an AGN for a given stellar mass, $M_\star$, and redshift, $z$ as a function of the specific black hole accretion rate, $\lambda$ [ergs/s/$\MSUN$]; also called the incidence rate distribution, $\mathcal{P}(\lambda\ |\ M_\star, z)$. The specific accretion rate, $\lambda$, of an AGN is the rate at which mass is accreting onto the supermassive black hole. Model C from \citetalias{aird2013primus} successfully predicts the XLF and its evolution at $0.2<z<1.0$. The specific accretion rate, $ \lambda$, is related to the X-ray luminosity,
\begin{equation}\label{eq:Lx2lambda}
    L_{X, \rm \ AGN}^{2 - 10\ \rm keV} = \frac{1}{25}{ \lambda  \times 1.26 \times 10^{38} \times 0.002 \ M_\star \ [\rm ergs\ s^{-1}] },
\end{equation}
where the $ 0.002 \ M_\star$ factor represents the mass of the black hole, $M_\bullet$, and assumes correlation between $M_\bullet$ and the mass of the bulge, $M_{\rm bulge}$~\citep{marconi2003relation}. Additionally, we also assume $M_\star \approx M_{\rm bulge}$~\citepalias{aird2013primus}. For the mean of our Full$_{\rm phot}$, $M_\star = 10^{10.7} \ \MSUN$ and at the median redshift, $\langle z \rangle = 0.14$, we obtain the incidence rate distribution, $\mathcal{P}(\lambda\ |\ L_{X},\ M_\star, z)$, as a function of $L_{X, \rm \ AGN}^{2 - 10\ \rm keV}$ using Eq.~\ref{eq:Lx2lambda}. To obtain the $0.5-2$ keV mean observed X-ray luminosity, which is required to compare with the estimate from this work, we further need to convert the incidence rate distribution from $L_{X, \rm \ AGN}^{2 - 10\ \rm keV}$ to $L_{X, \rm \ AGN}^{0.5 - 2\ \rm keV}$.

An important factor that comes into play when performing a conversion from $2-10$ keV (hard X-ray band; HXB) to the $0.5 - 2$ keV (soft X-ray band; SXB) luminosity is the intrinsic obscuration of the AGN.  Our estimate of $L_{X,\ \rm AGN}$ represents the contribution from the obscured AGNs and the observed unobscured Type 1 AGNs, the dominating component at the luminosity range under concern (see e.g.,~\citealt{hasinger2008absorption}). We used the \cite{Comparat2019agnmodel} empirical obscuration model to obtain the observed HXB-to-SXB luminosity conversion; they self-consistently build an obscuration model based on observation works \citep{Ricci2017rad, buchner2017galaxy, ueda2014toward, aird2015evolution, buchner2015obscuration}. The \cite{Comparat2019agnmodel} model is implemented on the \textsc{uchuu} simulations (introduced in Sec.~\ref{subsec:XRB}), and we obtain the HXB to SXB conversion as a function of $L_{X, \rm \ AGN}^{2 - 10\ \rm keV}$. Finally, we obtain the desired \citetalias{aird2013primus}-based $\mathcal{P}(\lambda\ |\ L_{X},\ M_\star, z)$ distribution as a function of the $L_{X, \rm \ AGN}^{0.5 - 2\ \rm keV}$. The expectation value is the obtained as follows: $\langle L_{X, \rm \ AGN}^{0.5 - 2\ \rm keV} \rangle = \int \mathcal{P}(\lambda\ |\ L_{X},\ M_\star, z) L_{X, \rm \ AGN}^{0.5 - 2\ \rm keV} d\lambda$.

An additional consideration is that the optical sample used for X-ray stacking in \citetalias{zhang2024hot} excludes objects classified as point sources in optical. This effectively excludes the optically bright quasars, where the point-like emission strongly dominates over the host galaxy contribution. An unsolved and open question is how such optical selection criteria for AGNs modify the X-ray luminosity distribution in X-rays, and addressing this is beyond the scope of this work. Nonetheless, we proceeded to compute a conservative X-ray luminosity threshold to account for this exclusion of optical quasars as follows. We converted the optical r-band luminosity distribution of the Full$_{\rm phot}$ galaxy sample to the $2-10$ keV luminosity distribution using a bolometric correction factor of $2.5$~\citep{collin2002quasars, duras2020universal, buchner2024genuine}. We used $10\times$ the mean of the HXB luminosity distribution as the threshold above which the object is classified as a bright point source in the optical LS DR9 catalog. This conservative limit excludes objects with $L_{X, \rm \ AGN}^{2 - 10\ \rm keV} > 8\times 10^{43}$ ergs/s. We adopted this cut in the $\mathcal{P}(\lambda\ |\ L_{X},\ M_\star, z)$ distribution as a function of the $L_{X, \rm \ AGN}^{2 - 10\ \rm keV}$ from \citetalias{aird2013primus}. After applying the obscuration model from \cite{Comparat2019agnmodel}, we obtained  
$\langle L_{X, \rm \ AGN}^{0.5 - 2\ \rm keV} \rangle = 2.7_{-2.0}^{+2.2}\times 10^{40}$ ergs/s. The sum of $L_{X, \rm \ AGN}$ computed here and $L_{X, \rm \ XRB}$ computed in the previous Sec.~\ref{subsec:XRB} is represented with the shaded green region in Fig.~\ref{fig:exclusion_models}.

The large error bars on our estimate of $L_{X,\rm \ AGN}$ using the methodology described here are due to the uncertainties in the empirical obscuration model and the uncertainties in the incidence rate distribution, which is poorly constrained for the low-redshift Universe. The estimates here can be further improved with future works that strengthen the connection between the low-luminosity X-ray AGN population and the host galaxy properties, proper knowledge mapping AGN selection functions from optical to X-ray luminosities, and better constrained obscuration models.

\begin{figure}

    \centering
    \includegraphics[width=\linewidth]{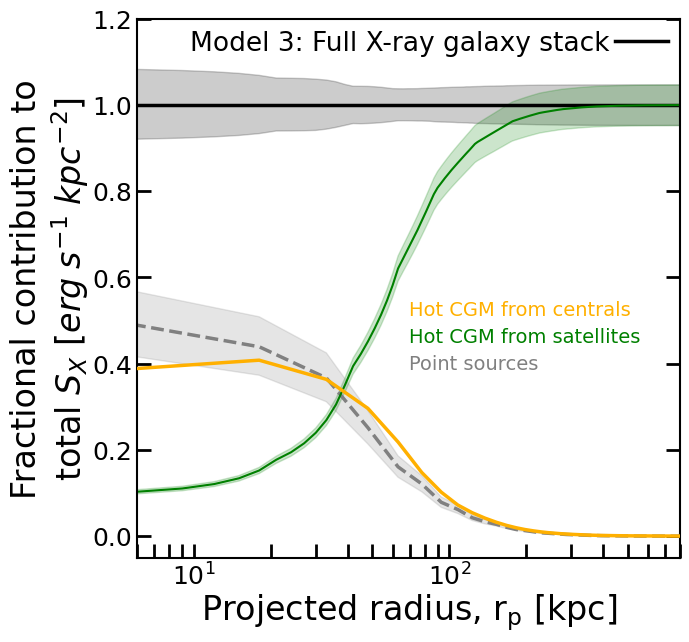}
    \centering
    \caption{Fractional contribution of centrals, satellites, and point sources to the total X-ray surface brightness profile of the hot CGM. The yellow line shows the contribution from central galaxies, the green line shows the events around satellites probing the hot gas of their more massive host halos, and the dashed gray line shows the X-ray events from unresolved and resolved point-like sources comprising AGNs and XRBs. The errors on the profiles were obtained from the posterior distributions of the MCMC fitting analysis.}
    \label{fig:frac_Sx}
\end{figure}

\subsection{Using Model $3$ to interpret the Full$_{\rm phot}$ data}
\label{subsec:model3}

In the light of the empirical estimates we obtain from Sec.~\ref{subsec:XRB} and \ref{subsec:AGN}, we compare the prediction for point source luminosities from our three forward models (based on the different halo distribution shown in Fig.~\ref{fig:halo-models}) with the empirically allowed range of point source luminosities as shown in Fig.~\ref{fig:exclusion_models}. This comparison favors model three, which has a mean $M_{\rm 200m} = 3.5\times 10^{12}\ \MSUN$, implying that the hot CGM component allows for a point source component with a luminosity that agrees with empirical estimates from the low-redshift Universe. We focus our results on model three for all the following discussions of the hot CGM. The results of fitting model $3$ to the X-ray surface brightness profile obtained by stacking on the Full$_{\rm phot}$ galaxies are shown in Fig.~\ref{fig:best-fit}, with the posterior distribution of the best-fit parameters shown in Fig.~\ref{fig:best-fit-posterior}.

\subsection{X-ray emission from the hot CGM}
\label{subsec:CGM}

The contribution of the hot CGM component from central galaxies is shown with the yellow line in Fig.~\ref{fig:best-fit} for the forward-model $3$. By integrating the area under the mean X-ray surface brightness profile from the central galaxy hot CGM component within $R_{500c}$, we obtain an X-ray luminosity, $L_{X, \ \rm CGM} =1.69_{-0.91}^{+0.28}\times 10^{40}$ ergs/s. We also show the residual plot of the percentage deviation of the data from our model, where the discrepancies are within $15\%$.

We show the fractional contribution of the various emission components in our forward model $3$ to the mean X-ray surface brightness profile upon stacking galaxies in Fig.~\ref{fig:frac_Sx}. We note that at mean redshifts of $0.14$ and the underlying halo mass distribution for model~$3$, the hot CGM is unresolved with an eROSITA-like PSF. Thus, at $\lesssim 40$ kpc, the hot CGM from central galaxies and the X-ray point sources emission from XRBs and AGNs each account for up to $40-50\%$ of the total X-ray emission budget, respectively.

\begin{figure}
    \centering
    \includegraphics[width=\linewidth]{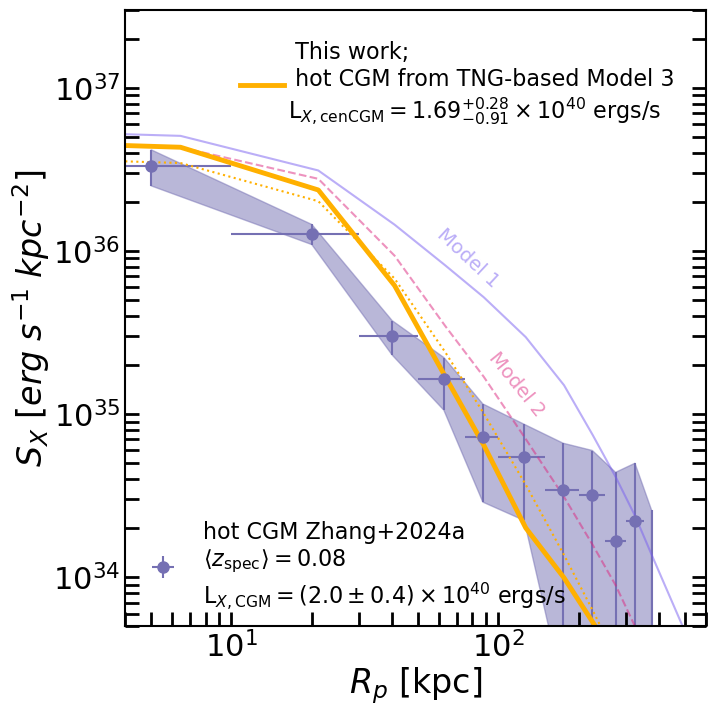}
    \centering
    \caption{Comparison of the hot gas CGM profile from our TNG-based model $3$ (solid yellow line) in this work with the hot CGM measurement from \citetalias{zhang2024hot} based on X-ray stacking at the optical positions of galaxies from the SDSS spectroscopic galaxy catalog (CEN sample). The CEN sample has a mean redshift $\langle z_{\rm spec}\rangle=0.08$, which is lower than that for the Full$_{\rm phot}$ galaxy sample, $\langle z_{\rm phot}\rangle = 0.14$, modeled in this work. Therefore, our models (solid yellow, dashed pink, and solid purple) are convolved with the eROSITA PSF representing $z=0.08$ to enable comparison with the CEN sample profile from \citetalias{zhang2024hot}. Our Full$_{\rm phot}$-based model~$3$ convolved with the $\langle z_{\rm phot}\rangle = 0.14$ PSF in shown by the dotted yellow line. For reference, we also show the other two models, $1$ and $2$, with different underlying halo distributions (see text in Sec.~\ref{subsec:mock}), which we excluded in this work (see Fig.~\ref{fig:exclusion_models}) as the hot CGM component did not allow for a point source component with a luminosity consistent with empirical estimates from the low-redshift Universe. The model~$3$ from this work is in good agreement with the CEN sample hot CGM profile from \citetalias{zhang2024hot}.}
    \label{fig:compareZ24}
\end{figure}

We compare our results with the other hot CGM measurements presented in \citetalias{zhang2024hot}, based on a different optical galaxy catalog, namely from the SDSS spectroscopic survey. Given the spectroscopic optical information, the galaxy sample is classified into centrals and satellites~\citep{tinker2021self}, which makes it possible to empirically model the hot CGM profile from other contaminating effects (point sources and satellites). They selected $30, 825$ central galaxies with spectroscopic redshifts $<0.2$ and MW-like stellar masses of $10.5 < \log(M_\star/\MSUN) < 11$. In \citetalias{zhang2024hot}, this SDSS-based spectroscopic sample is called the CEN sample. The resulting profile, as shown by the data points in Fig.~\ref{fig:compareZ24}, is compared with the hot CGM component (model $3$) we obtain in this work (solid yellow line). Our TNG-based forward model of the hot CGM prediction is in excellent agreement with the hot CGM measurement from \citetalias{zhang2024hot} at $\gtrsim 60$~kpc. At the inner radii ($\lesssim 60$~kpc), our TNG-based model~$3$ overpredicts the X-ray emission. We note that the halo mass distributions of the two samples have similar mean values, where the mean $M_{\rm 200m} = 3.5\times 10^{12}\ \MSUN$ for our forward model $3$ and the mean $M_{\rm 200m}^{\rm CEN} \sim 3 \times 10^{12}\ \MSUN$. However, the median $M_{\rm 200m}^{\rm CEN} \sim 1.3 \times 10^{12}\ \MSUN$, highlighting the spread in the halo mass distribution. This result further emphasizes the importance of the underlying halo mass distribution and the impact of the halo mass scatter introduced in stellar-mass selections when comparing hot CGM profiles across different observations and simulation-based models. For reference, we also show the forward models $1$ and $2$, which we exclude because their hot CGM component does not allow for a permissible contribution of point source luminosity in the X-ray galaxy stack (see Sec.~\ref{subsec:model3}). In addition to this shortcoming, we find that models $1$ and $2$ are discrepant with the CEN sample measurement of the hot CGM, further favoring model $3$.

\subsection{X-ray emission from satellite's host halos}
\label{subsec:sat}

At larger radii ($\gtrsim 40$ kpc), the contribution from the emission around satellites dominates the total X-ray signal, thereby explaining the flattening observed in the measurements, {as also found by \cite{Comparat2025cross}}. This emission originates from the massive host halos in which the satellite galaxies reside; we find the mean host halo mass of these satellites to be $M_{\rm 200m} \sim 10^{14}\ \MSUN$. {We emphasize that this component is not intended to probe the intrinsic CGM of the satellites themselves, but instead captures the background X-ray emission associated with their host halos. Since the Full$_{\rm phot}$ galaxy catalog includes both central and satellite galaxies without classification, any observational stacking analysis based on such samples unavoidably includes this contribution.} We quantify the X-ray luminosity from satellite host halos as $L_{X,\ \rm SAT} = 3_{-2}^{+3} \times 10^{41}$~erg/s, obtained by integrating the area under the predicted surface brightness profile ${S}_{\rm X,\ sat}$. {Interestingly, \cite{zhang2018emission} also find an inflection point at $\sim 50$ kpc due to satellite's host halos in the stacked H$\alpha+$N[\textsc{ii}] radial emission profile (see Fig.~8 therein), where they trace the cool component of the CGM. The findings from \cite{zhang2018emission} on the cool CGM is complementary to our results presented here on forward-modeling the hot CGM probed by X-ray stacking.}

Upon fitting our forward model to the Full$_{\rm phot}$ stack, we introduced a renormalization parameter, $\mathcal{N}_{\rm sat}$, which rescales the ${S}_{\rm X,\ sat}$ contribution to match the observations. This scaling was necessary due to the known overprediction of hot gas fractions in TNG300 for halos above $10^{13.5}\ \MSUN$ \citep{popesso2024hot}, leading to an overly bright ${S}_{\rm X,\ sat}$ component (see Sec.~\ref{subsec:fmodel_hot}). {Importantly, the shape of this satellite-related emission component is independent of the satellite fraction, as demonstrated in Fig.~9 in \cite{shreeram2025quantifying}.} From our fitting analysis for model 3, we find that the TNG-based fiducial ${S}_{\rm X,\ sat}$ normalization of the hot CGM  must be rescaled by $0.15$ to provide an observationally consistent contribution for the Full$_{\rm phot}$ galaxy catalog based X-ray stack. {We highlight that this renormalization does not affect the hot CGM contribution from centrals, which remains unchanged. Therefore, $\mathcal{N}_{\rm sat}$ effectively provides a quantitative correction factor, illustrating how the overabundance of hot gas in massive halos biases the observable MW-mass stacked X-ray profiles. We conclude that the inclusion and forward-modeling of the satellite's host halo component is required to physically interpret the Full$_{\rm phot}$ data.}

\section{Summary}
\label{sec:summary}

In this work, we have forward-modeled the measurements of the X-ray surface brightness profiles obtained by stacking at the optical galaxy positions of the LS DR9 photometric (Full$_{\rm phot}$) galaxy catalog, reported by \citetalias{zhang2024hot}. We retrieved the contribution of the hot CGM from central galaxies from that of point sources and satellite galaxies. Our hot CGM forward model is based on TNG300 hydrodynamical simulations. The main results from this work are summarized as follows:
\begin{enumerate}
    \item  We tested the impact of the underlying halo mass distribution on the TNG-based prediction for corresponding X-ray surface brightness profiles. We did so by generating multiple mock galaxy catalogs using LC-TNG300, matched in stellar mass and redshift to the Full$_{\rm phot}$ galaxy catalog; however, with different underlying halo mass distributions (Sec.~\ref{subsec:mock}; Fig.~\ref{fig:halo-models}).  Namely, the three models obtained by varying the halo mass distribution are as follows: model $1$ leaves the LC-TNG300 halo mass distribution unchanged, whereas models $2$ and $3$ change the underlying halo distribution by excluding the $10\%$ and  $30\%$ most massive halos from the original LC-TNG300 halo mass distribution, respectively. We show that the shape and normalization of the hot CGM X-ray surface brightness profiles are impacted by varying the halo mass distributions, whereby discarding the most massive halos from the underlying halo distribution results in steeper profiles with lower normalization. More precisely, we find that a factor $\sim 2\times$ increase in the mean value of the underlying halo mass distribution results in an $\sim 4\times$ increase in the X-ray luminosity from the hot CGM. \\

    \item We fit the stacked X-ray radial surface brightness profile by eROSITA around MW-mass galaxies from \cite{zhang2024hot} with our forward models. Our model contains two emitting components (Eq.~\ref{eq:full_Sx}-\ref{eq:hot_gas}): hot gas (around central galaxies and around satellite galaxies hosted by more massive halos) and X-ray point sources (XRBs and AGNs). For three forward models, we computed the X-ray luminosity from point sources, $L_{X,\ \rm PS}$, and CGM (see results in Tab.~\ref{tab:best-fit-params0}). Using the empirical estimates for the expected luminosity from XRBs (Sec.~\ref{subsec:XRB}) and AGNs (Sec.~\ref{subsec:AGN}) for MW-mass galaxies, we put constraints on the permissible values of the $L_{X,\ \rm PS}$ contribution to the X-ray stack (see Fig.~\ref{fig:halo-models}). This analysis favors model three, which has a mean $M_{\rm 200m} = 3.5\times 10^{12}\ \MSUN$, implying that the hot CGM component allows for a point source component with a mean AGN luminosity that agrees with empirical estimates from the low-redshift Universe. We focus our results on model three for all the following discussions of the hot CGM.\\

    \item By integrating the area under the mean X-ray surface brightness profile from the central galaxy hot CGM component within $R_{500c}$, we obtain an X-ray luminosity of $L_{X, \ \rm CGM} =1.69_{-0.91}^{+0.28}\times 10^{40}$ ergs/s. We also show the residual plot of the percentage deviation of the data from our model, where the discrepancies within the $50-105$ kpc range are within $15\%$ (Fig.~\ref{fig:best-fit}). We find that at $\lesssim 40$ kpc, the hot CGM from central galaxies and the X-ray point sources emission from XRBs and AGNs each account for $40-50\%$ of the total X-ray emission budget, respectively (Fig.~\ref{fig:frac_Sx}). At larger radii $>40$ kpc, the contribution from the emission around satellites dominates the total X-ray emission, thereby explaining the overall flattening in the measurements.\\

    \item  We compared our results with the other hot CGM measurements presented in \citetalias{zhang2024hot}, based on a different optical galaxy catalog, namely from the SDSS spectroscopic survey (see comparison in Fig.~\ref{fig:compareZ24}). Our TNG-based forward model of the hot CGM prediction broadly agrees with the hot CGM measurement from \citetalias{zhang2024hot}. The $L_{X, \ \rm CGM}$ measured between the two works is consistent. We note that the halo mass distributions of the two samples are similar mean values, where the mean $M_{\rm 200m} = 3.5\times 10^{12}\ \MSUN$ for our forward model $3$ and the mean $M_{\rm 200m}^{\rm CEN} \sim 3 \times 10^{12}\ \MSUN$. This result further emphasizes the importance of the underlying halo mass distribution when comparing hot CGM profiles across different observations and simulation-based models.  \\
        
\end{enumerate}

This work provides a novel technique to constrain the mean AGN X-ray luminosity of a galaxy sample jointly with the radial hot CGM gas distribution within the halo using the X-ray hot CGM (stacking) measurements as a new benchmark. Alongside the progress in our understanding of how various stellar and AGN feedback prescriptions impact the hot CGM's properties~\citep{lau2024x, medlock2025quantifying}, here we emphasize another vital ingredient when comparing simulations with X-ray observations: the sensitivity of the X-ray CGM properties to the underlying halo mass distribution, stellar mass, and redshift. One of the outstanding challenges in the current paradigm of galaxy formation and evolution models implemented in hydrodynamical simulations is jointly constraining the microscopic scales (e.g., subgrid model physics) and their impact on the diffuse gas within the halo~\citep{crain2023hydrodynamical}. Future work implementing the data-comparison strategy developed here on other state-of-the-art simulations, such as EAGLE~\citep{Crain2015eagle, Schaye2015eagle}, FLAMINGO~\citep{schaye2023flamingo}, Magneticum~\citep{Dolag2005magneticum, Beck2016magneticum}, and SIMBA~\citep{dave2019simba}, will provide observationally motivated ranges on the allowed X-ray AGN luminosity for the MW-mass scales. Comparing the AGN X-ray luminosity predictions retrieved from the methodology developed here (informed by hot CGM X-ray observations) with those predicted by the simulation itself will provide a new ground for recalibrating and improving the current landscape of sub-grid AGN modes (see e.g.,~\citealt{alexander2012drives} for a review).
Additionally, future X-ray missions on the observation side, such as Athena~\citep{nandra2013hot}, AXIS~\citep{mushotzky2019advanced}, and HUBS~\citep{cui2020hubs}, will push our current detection limits to resolve the hot CGM at higher redshifts in X-rays. This would further our understanding of how observations compare to the spatially resolved hot gas distribution at MW-mass scales in simulations.

\begin{acknowledgements}
SS would like to thank Fulvio Ferlito and Matteo Guardiani for all the useful scientific discussions. G.P. acknowledges financial support from the European Research Council (ERC) under the European Union's Horizon 2020 research and innovation program "Hot Milk" (grant agreement No. 865637) and support from Bando per il Finanziamento della Ricerca Fondamentale 2022 dell'Istituto Nazionale di Astrofisica (INAF): GO Large program and from the Framework per l'Attrazione e il Rafforzamento delle Eccellenze (FARE) per la ricerca in Italia (R20L5S39T9).\\

Computations were performed on the HPC system Raven at the Max Planck Computing and Data Facility. We acknowledge the project support by the Max Planck Computing and Data Facility.
\end{acknowledgements}

\bibliographystyle{aa} 
\bibliography{biblio}

@article{biffi2018origin,
  title={The origin of ICM enrichment in the outskirts of present-day galaxy clusters from cosmological hydrodynamical simulations},
  author={Biffi, Veronica and Planelles, Susana and Borgani, Stefano and Rasia, Elena and Murante, Giuseppe and Fabjan, Dunja and Gaspari, Massimo},
  journal={MNRAS},
  volume={476},
  number={2},
  pages={2689--2703},
  year={2018},
  publisher={Oxford University Press}
}

@article{comparat2022erosita,
  title={The eROSITA Final Equatorial Depth Survey (eFEDS)-X-ray emission around star-forming and quiescent galaxies at 0.05< z< 0.3},
  author={Comparat, Johan and Truong, Nhut and Merloni, Andrea and Pillepich, Annalisa and Ponti, Gabriele and Driver, Simon and Bellstedt, Sabine and Liske, Joe and Aird, James and Br{\"u}ggen, Marcus and others},
  journal={A\&A},
  volume={666},
  pages={A156},
  year={2022},
  publisher={EDP Sciences}
}

@ARTICLE{ComparatEckertFinoguenov_2020,
       author = {{Comparat}, Johan and {Eckert}, Dominique and {Finoguenov}, Alexis and {Schmidt}, Robert and {Sanders}, Jeremy S. and {Nagai}, Daisuke and {Lau}, Erwin T. and {K{\"a}} and {fer}, Florian and {Pacaud}, Florian and et al.},
        title = "{Full-sky photon simulation of clusters and active galactic nuclei in the soft X-rays for eROSITA}",
      journal = {OJAp},
         year = 2020,
        month = dec,
       volume = {3},
       number = {1},
          eid = {13},
        pages = {13},
          doi = {10.21105/astro.2008.08404},
archivePrefix = {arXiv},
       eprint = {2008.08404},
 primaryClass = {astro-ph.CO},
       adsurl = {https://ui.adsabs.harvard.edu/abs/2020OJAp....3E..13C}
}

@article{Zhang:2022un,
	adsurl = {https://ui.adsabs.harvard.edu/abs/2022A&A...663A..85Z},
	archiveprefix = {arXiv},
	author = {{Zhang}, Ziwen and {Wang}, Huiyuan and {Luo}, Wentao and {Zhang}, Jun and {Mo}, Houjun J. and {Jing}, YiPeng and {Yang}, Xiaohu and {Li}, Hao},
	doi = {10.1051/0004-6361/202142866},
	eid = {A85},
	eprint = {2112.04777},
	journal = {\aap},
	month = jul,
	pages = {A85},
	primaryclass = {astro-ph.GA},
	title = {{Massive star-forming galaxies have converted most of their halo gas into stars}},
	volume = {663},
	year = 2022
	}

@article{pillepich2018first,
	author = {Pillepich, Annalisa and Nelson, Dylan and Hernquist, Lars and Springel, Volker and Pakmor, R{\"u}diger and Torrey, Paul and Weinberger, Rainer and Genel, Shy and Naiman, Jill P and Marinacci, Federico and others},
	journal = {MNRAS},
	number = {1},
	pages = {648--675},
	publisher = {Oxford University Press},
	title = {First results from the IllustrisTNG simulations: the stellar mass content of groups and clusters of galaxies},
	volume = {475},
	year = {2018}}

@article{nelson2015illustris,
	author = {Nelson, Dylan and Pillepich, Annalisa and Genel, Shy and Vogelsberger, Mark and Springel, Volker and Torrey, Paul and Rodriguez-Gomez, Vicente and Sijacki, Debora and Snyder, Gregory F and Griffen, Brendan and others},
	journal = {A\&C},
	pages = {12--37},
	publisher = {Elsevier},
	title = {The illustris simulation: Public data release},
	volume = {13},
	year = {2015}}

@article{Biffi:2018wj,
	adsurl = {https://ui.adsabs.harvard.edu/abs/2018MNRAS.481.2213B},
	archiveprefix = {arXiv},
	author = {{Biffi}, V. and {Dolag}, K. and {Merloni}, A.},
	doi = {10.1093/mnras/sty2436},
	eprint = {1804.01096},
	journal = {\mnras},
	month = dec,
	number = {2},
	pages = {2213-2227},
	primaryclass = {astro-ph.CO},
	title = {{AGN contamination of galaxy-cluster thermal X-ray emission: predictions for eRosita from cosmological simulations}},
	volume = {481},
	year = 2018}

@article{Biffi:2013uh,
	adsurl = {https://ui.adsabs.harvard.edu/abs/2013MNRAS.428.1395B},
	archiveprefix = {arXiv},
	author = {{Biffi}, V. and {Dolag}, K. and {B{\"o}hringer}, H.},
	doi = {10.1093/mnras/sts120},
	eprint = {1210.4158},
	journal = {\mnras},
	month = jan,
	number = {2},
	pages = {1395-1409},
	primaryclass = {astro-ph.CO},
	title = {{Investigating the velocity structure and X-ray observable properties of simulated galaxy clusters with PHOX}},
	volume = {428},
	year = 2013}

@article{Chadayammuri:2022us,
	adsurl = {https://ui.adsabs.harvard.edu/abs/2022ApJ...936L..15C},
	archiveprefix = {arXiv},
	author = {{Chadayammuri}, Urmila and {Bogd{\'a}n}, {\'A}kos and {Oppenheimer}, Benjamin D. and {Kraft}, Ralph P. and {Forman}, William R. and {Jones}, Christine},
	doi = {10.3847/2041-8213/ac8936},
	eid = {L15},
	eprint = {2203.01356},
	journal = {\apjl},
	month = sep,
	number = {1},
	pages = {L15},
	primaryclass = {astro-ph.GA},
	title = {{Testing Galaxy Feedback Models with Resolved X-Ray Profiles of the Hot Circumgalactic Medium}},
	volume = {936},
	year = 2022
	}

@article{Vladutescu-Zopp:2023vo,
	adsurl = {https://ui.adsabs.harvard.edu/abs/2023A&A...669A..34V},
	archiveprefix = {arXiv},
	author = {{Vladutescu-Zopp}, S. and {Biffi}, V. and {Dolag}, K.},
	doi = {10.1051/0004-6361/202244726},
	eid = {A34},
	eprint = {2208.04975},
	journal = {\aap},
	month = jan,
	pages = {A34},
	primaryclass = {astro-ph.HE},
	title = {{Decomposition of galactic X-ray emission with PHOX. Contributions from hot gas and X-ray binaries}},
	volume = {669},
	year = 2023}

@article{zhang2024hot,
  title={The hot circumgalactic medium in the eROSITA All-Sky Survey-I. X-ray surface brightness profiles},
  author={Zhang, Yi and Comparat, Johan and Ponti, Gabriele and Merloni, Andrea and Nandra, Kirpal and Haberl, Frank and Locatelli, Nicola and Zhang, Xiaoyuan and Sanders, Jeremy and Zheng, Xueying and others},
  journal={A\&A},
  volume={690},
  pages={A267},
  year={2024},
  publisher={EDP Sciences}
}

@article{zuhone2016pyxsim,
  title={pyXSIM: Synthetic X-ray observations generator},
  author={ZuHone, John A and Hallman, Eric J},
  journal={Astrophysics Source Code Library},
  pages={ascl--1608},
  year={2016}
}

@article{carlson2010embedding,
  title={Embedding realistic surveys in simulations through volume remapping},
  author={Carlson, Jordan and White, Martin},
  journal={ApJ Supplement Series},
  volume={190},
  number={2},
  pages={311},
  year={2010},
  publisher={IOP Publishing}
}

@article{ade2016planck,
  title={Planck 2015 results-xiii. cosmological parameters},
  author={Ade, Peter AR and Aghanim, Nabila and Arnaud, M and Ashdown, Mark and Aumont, J and Baccigalupi, Carlo and Banday, AJ and Barreiro, RB and Bartlett, JG and Bartolo, Nicola and others},
  journal={A\&A},
  volume={594},
  pages={A13},
  year={2016},
  publisher={EDP sciences}
}

@article{smith2001collisional,
  title={Collisional plasma models with APEC/APED: emission-line diagnostics of hydrogen-like and helium-like ions},
  author={Smith, Randall K and Brickhouse, Nancy S and Liedahl, Duane A and Raymond, John C},
  journal={ApJ},
  volume={556},
  number={2},
  pages={L91},
  year={2001},
  publisher={IOP Publishing}
}

@article{naiman2018first,
  title={First results from the IllustrisTNG simulations: A tale of two elements--chemical evolution of magnesium and europium},
  author={Naiman, Jill P and Pillepich, Annalisa and Springel, Volker and Ramirez-Ruiz, Enrico and Torrey, Paul and Vogelsberger, Mark and Pakmor, R{\"u}diger and Nelson, Dylan and Marinacci, Federico and Hernquist, Lars and others},
  journal={MNRAS},
  volume={477},
  number={1},
  pages={1206--1224},
  year={2018},
  publisher={Oxford University Press}
}

@article{nelson2019illustristng,
  title={The IllustrisTNG simulations: public data release},
  author={Nelson, Dylan and Springel, Volker and Pillepich, Annalisa and Rodriguez-Gomez, Vicente and Torrey, Paul and Genel, Shy and Vogelsberger, Mark and Pakmor, Ruediger and Marinacci, Federico and Weinberger, Rainer and others},
  journal={CompAC},
  volume={6},
  pages={1--29},
  year={2019},
  publisher={Springer}
}

@article{marinacci2018first,
  title={First results from the IllustrisTNG simulations: radio haloes and magnetic fields},
  author={Marinacci, Federico and Vogelsberger, Mark and Pakmor, R{\"u}diger and Torrey, Paul and Springel, Volker and Hernquist, Lars and Nelson, Dylan and Weinberger, Rainer and Pillepich, Annalisa and Naiman, Jill and others},
  journal={MNRAS},
  volume={480},
  number={4},
  pages={5113--5139},
  year={2018},
  publisher={Oxford University Press}
}

@article{springel2018first,
  title={First results from the IllustrisTNG simulations: matter and galaxy clustering},
  author={Springel, Volker and Pakmor, R{\"u}diger and Pillepich, Annalisa and Weinberger, Rainer and Nelson, Dylan and Hernquist, Lars and Vogelsberger, Mark and Genel, Shy and Torrey, Paul and Marinacci, Federico and others},
  journal={MNRAS},
  volume={475},
  number={1},
  pages={676--698},
  year={2018},
  publisher={Oxford University Press}
}

@article{springel2001populating,
  title={Populating a cluster of galaxies--I. Results at z= 0},
  author={Springel, Volker and White, Simon DM and Tormen, Giuseppe and Kauffmann, Guinevere},
  journal={MNRAS},
  volume={328},
  number={3},
  pages={726--750},
  year={2001},
  publisher={Blackwell Science Ltd Oxford, UK}
}

@article{dolag2009substructures,
  title={Substructures in hydrodynamical cluster simulations},
  author={Dolag, K and Borgani, Stefano and Murante, G and Springel, V},
  journal={MNRAS},
  volume={399},
  number={2},
  pages={497--514},
  year={2009},
  publisher={Blackwell Publishing Ltd Oxford, UK}
}

@article{foster2012updated,
  title={Updated atomic data and calculations for X-ray spectroscopy},
  author={Foster, AR and Ji, L and Smith, RK and Brickhouse, NS},
  journal={ApJ},
  volume={756},
  number={2},
  pages={128},
  year={2012},
  publisher={IOP Publishing}
}

@article{merloni2024srg,
  title={The SRG/eROSITA all-sky survey-First X-ray catalogues and data release of the western Galactic hemisphere},
  author={Merloni, A and Lamer, G and Liu, T and Ramos-Ceja, ME and Brunner, H and Bulbul, E and Dennerl, K and Doroshenko, V and Freyberg, MJ and Friedrich, S and others},
  journal={A\&A},
  volume={682},
  pages={A34},
  year={2024},
  publisher={EDP Sciences}
}

@article{dey2019overview,
  title={Overview of the DESI legacy imaging surveys},
  author={Dey, Arjun and Schlegel, David J and Lang, Dustin and Blum, Robert and Burleigh, Kaylan and Fan, Xiaohui and Findlay, Joseph R and Finkbeiner, Doug and Herrera, David and Juneau, St{\'e}phanie and others},
  journal={ApJ},
  volume={157},
  number={5},
  pages={168},
  year={2019},
  publisher={IOP Publishing}
}

@article{zou2019photometric,
  title={Photometric Redshifts and Stellar Masses for Galaxies from the DESI Legacy Imaging Surveys},
  author={Zou, Hu and Gao, Jinghua and Zhou, Xu and Kong, Xu},
  journal={ApJ Supplement Series},
  volume={242},
  number={1},
  pages={8},
  year={2019},
  publisher={IOP Publishing}
}

@article{zou2022photometric,
  title={Photometric Redshifts and Galaxy Clusters for DES DR2, DESI DR9, and HSC-SSP PDR3 Data},
  author={Zou, Hu and Sui, Jipeng and Xue, Suijian and Zhou, Xu and Ma, Jun and Zhou, Zhimin and Nie, Jundan and Zhang, Tianmeng and Feng, Lu and Shen, Zhixia and others},
  journal={RAA},
  volume={22},
  number={6},
  pages={065001},
  year={2022},
  publisher={IOP Publishing}
}

@article{bekhti2016hi4pi,
  title={HI4PI: a full-sky H i survey based on EBHIS and GASS},
  author={Bekhti, N Ben and Fl{\"o}er, L and Keller, R and Kerp, J and Lenz, D and Winkel, B and Bailin, J and Calabretta, MR and Dedes, L and Ford, HA and others},
  journal={A\&A},
  volume={594},
  pages={A116},
  year={2016},
  publisher={EDP Sciences}
}

@article{strauss2002spectroscopic,
  title={Spectroscopic target selection in the Sloan Digital Sky Survey: the main galaxy sample},
  author={Strauss, Michael A and Weinberg, David H and Lupton, Robert H and Narayanan, Vijay K and Annis, James and Bernardi, Mariangela and Blanton, Michael and Burles, Scott and Connolly, AJ and Dalcanton, Julianne and others},
  journal={ApJ},
  volume={124},
  number={3},
  pages={1810},
  year={2002},
  publisher={IOP Publishing}
}

@article{tinker2021self,
  title={A Self-Calibrating Halo-Based Group Finder: Application to SDSS},
  author={Tinker, Jeremy L},
  journal={ApJ},
  volume={923},
  number={2},
  pages={154},
  year={2021},
  publisher={IOP Publishing}
}

@article{abazajian2009seventh,
  title={The seventh data release of the Sloan Digital Sky Survey},
  author={Abazajian, Kevork N and Adelman-McCarthy, Jennifer K and Ag{\"u}eros, Marcel A and Allam, Sahar S and Prieto, Carlos Allende and An, Deokkeun and Anderson, Kurt SJ and Anderson, Scott F and Annis, James and Bahcall, Neta A and others},
  journal={ApJ Supplement Series},
  volume={182},
  number={2},
  pages={543},
  year={2009},
  publisher={IOP Publishing}
}

@article{anderson2015unifying,
  title={Unifying X-ray scaling relations from galaxies to clusters},
  author={Anderson, Michael E and Gaspari, Massimo and White, Simon DM and Wang, Wenting and Dai, Xinyu},
  journal={MNRAS},
  volume={449},
  number={4},
  pages={3806--3826},
  year={2015},
  publisher={The Royal Astronomical Society}
}

@article{oppenheimer2020eagle,
  title={EAGLE and Illustris-TNG predictions for resolved eROSITA X-ray observations of the circumgalactic medium around normal galaxies},
  author={Oppenheimer, Benjamin D and Bogd{\'a}n, {\'A}kos and Crain, Robert A and ZuHone, John A and Forman, William R and Schaye, Joop and Wijers, Nastasha A and Davies, Jonathan J and Jones, Christine and Kraft, Ralph P and others},
  journal={ApJ Letters},
  volume={893},
  number={1},
  pages={L24},
  year={2020},
  publisher={IOP Publishing}
}

@article{tumlinson2017circumgalactic,
  title={The circumgalactic medium},
  author={Tumlinson, Jason and Peeples, Molly S and Werk, Jessica K},
  journal={ARAA},
  volume={55},
  pages={389--432},
  year={2017},
  publisher={Annual Reviews}
}

@ARTICLE{Wechsler2018halo,
       author = {{Wechsler}, Risa H. and {Tinker}, Jeremy L.},
        title = "{The Connection Between Galaxies and Their Dark Matter Halos}",
      journal = {\araa},
         year = 2018,
        month = sep,
       volume = {56},
        pages = {435-487},
          doi = {10.1146/annurev-astro-081817-051756},
archivePrefix = {arXiv},
       eprint = {1804.03097},
 primaryClass = {astro-ph.GA},
       adsurl = {https://ui.adsabs.harvard.edu/abs/2018ARA&A..56..435W}
}

@article{faucher2023key,
  title={Key physical processes in the circumgalactic medium},
  author={Faucher-Gigu{\`e}re, Claude-Andr{\'e} and Oh, S Peng},
  journal={ARAA},
  volume={61},
  pages={131--195},
  year={2023},
  publisher={Annual Reviews}
}

@ARTICLE{Wijers2020eagle,
       author = {{Wijers}, Nastasha A. and {Schaye}, Joop and {Oppenheimer}, Benjamin D.},
        title = "{The warm-hot circumgalactic medium around EAGLE-simulation galaxies and its detection prospects with X-ray and UV line absorption}",
      journal = {\mnras},
         year = 2020,
        month = oct,
       volume = {498},
       number = {1},
        pages = {574-598},
          doi = {10.1093/mnras/staa2456},
archivePrefix = {arXiv},
       eprint = {2004.05171},
 primaryClass = {astro-ph.GA},
       adsurl = {https://ui.adsabs.harvard.edu/abs/2020MNRAS.498..574W}
}

@ARTICLE{Bertone2010oviii,
       author = {{Bertone}, Serena and {Schaye}, Joop and {Dalla Vecchia}, Claudio and {Booth}, C.~M. and {Theuns}, Tom and {Wiersma}, Robert P.~C.},
        title = "{Metal-line emission from the warm-hot intergalactic medium - I. Soft X-rays}",
      journal = {\mnras},
         year = 2010,
        month = sep,
       volume = {407},
       number = {1},
        pages = {544-566},
          doi = {10.1111/j.1365-2966.2010.16932.x},
archivePrefix = {arXiv},
       eprint = {0910.5723},
 primaryClass = {astro-ph.CO},
       adsurl = {https://ui.adsabs.harvard.edu/abs/2010MNRAS.407..544B}
}

@ARTICLE{van2013owl,
       author = {{van de Voort}, Freeke},
        title = "{Soft X-ray and ultraviolet metal-line emission from the gas around galaxies}",
      journal = {\mnras},
         year = 2013,
        month = apr,
       volume = {430},
       number = {4},
        pages = {2688-2702},
          doi = {10.1093/mnras/stt115},
archivePrefix = {arXiv},
       eprint = {1207.5512},
 primaryClass = {astro-ph.CO},
       adsurl = {https://ui.adsabs.harvard.edu/abs/2013MNRAS.430.2688V}
}

@article{galeazzi2007xmm, title={XMM-Newton observations of the diffuse X-ray background}, author={Galeazzi, M and Gupta, Anjali and Covey, K and Ursino, E}, journal={ApJ}, volume={658}, number={2}, pages={1081}, year={2007}, publisher={IOP Publishing} }

@article{bogdan2023circumgalactic,
  title={Circumgalactic Medium on the Largest Scales: Detecting X-Ray Absorption Lines with Large-area Microcalorimeters},
  author={Bogd{\'a}n, {\'A}kos and Khabibullin, Ildar and Kov{\'a}cs, Orsolya E and Schellenberger, Gerrit and ZuHone, John and Burchett, Joseph N and Dolag, Klaus and Churazov, Eugene and Forman, William R and Jones, Christine and others},
  journal={ApJ},
  volume={953},
  number={1},
  pages={42},
  year={2023},
  publisher={IOP Publishing},
url = {https://iopscience.iop.org/article/10.3847/1538-4357/acdeec/meta}
}

@article{mathur2023probing,
  title={Probing the hot circumgalactic medium of external galaxies in X-ray absorption II: a luminous spiral galaxy at z≈ 0.225},
  author={Mathur, Smita and Das, Sanskriti and Gupta, Anjali and Krongold, Yair},
  journal={MNRAS: Letters},
  volume={525},
  number={1},
  pages={L11--L16},
  year={2023},
  publisher={Oxford University Press},
url={https://academic.oup.com/mnrasl/article/525/1/L11/7205507}
}

@article{bhattacharyya2023hot, title={The Hot Circumgalactic Medium of the Milky Way: New Insights from XMM-Newton Observations}, author={Bhattacharyya, Joy and Das, Sanskriti and Gupta, Anjali and Mathur, Smita and Krongold, Yair}, journal={ApJ}, volume={952}, number={1}, pages={41}, year={2023}, publisher={IOP Publishing} }

@article{li2017circum,
  title={The circum-galactic medium of massive spirals. II. Probing the nature of hot gaseous halo around the most massive isolated spiral galaxies},
  author={Li, Jiang-Tao and Bregman, Joel N and Wang, Q Daniel and Crain, Robert A and Anderson, Michael E and Zhang, Shangjia},
  journal={ApJ Supplement Series},
  volume={233},
  number={2},
  pages={20},
  year={2017},
  publisher={IOP Publishing},
url={https://iopscience.iop.org/article/10.3847/1538-4365/aa96fc}
}

@article{bogdan2013detection,
  title={Detection of a luminous hot X-ray corona around the massive spiral galaxy NGC 266},
  author={Bogd{\'a}n, {\'A}kos and Forman, William R and Kraft, Ralph P and Jones, Christine},
  journal={ApJ},
  volume={772},
  number={2},
  pages={98},
  year={2013},
  publisher={IOP Publishing},
url={https://iopscience.iop.org/article/10.1088/0004-637X/772/2/97}
}

@article{bogdan2013hot,
  title={Hot X-ray coronae around massive spiral galaxies: a unique probe of structure formation models},
  author={Bogd{\'a}n, {\'A}kos and Forman, William R and Vogelsberger, Mark and Bourdin, Herv{\'e} and Sijacki, Debora and Mazzotta, Pasquale and Kraft, Ralph P and Jones, Christine and Gilfanov, Marat and Churazov, Eugene and others},
  journal={ApJ},
  volume={772},
  number={2},
  pages={97},
  year={2013},
  publisher={IOP Publishing},
url={https://iopscience.iop.org/article/10.1088/0004-637X/772/2/97}
}

@article{bogdan2017probing,
  title={Probing the hot X-ray corona around the massive spiral galaxy, NGC 6753, using deep XMM-Newton observations},
  author={Bogd{\'a}n, {\'A}kos and Bourdin, Herv{\'e} and Forman, William R and Kraft, Ralph P and Vogelsberger, Mark and Hernquist, Lars and Springel, Volker},
  journal={ApJ},
  volume={850},
  number={1},
  pages={98},
  year={2017},
  publisher={IOP Publishing}
}

@article{anderson2016deep,
  title={A deep XMM--Newton study of the hot gaseous halo around NGC 1961},
  author={Anderson, Michael E and Churazov, Eugene and Bregman, Joel N},
  journal={MNRAS},
  volume={455},
  number={1},
  pages={227--243},
  year={2016},
  publisher={Oxford University Press}
}

@article{das2019evidence,
  title={Evidence for a Massive Warm--Hot Circumgalactic Medium around NGC 3221},
  author={Das, Sanskriti and Mathur, Smita and Gupta, Anjali and Nicastro, Fabrizio and Krongold, Yair and Null, Cody},
  journal={ApJ},
  volume={885},
  number={2},
  pages={108},
  year={2019},
  publisher={IOP Publishing}
}

@ARTICLE{rohr2024sat,
       author = {{Rohr}, Eric and {Pillepich}, Annalisa and {Nelson}, Dylan and {Ayromlou}, Mohammadreza and {Zinger}, Elad},
        title = "{The hot circumgalactic media of massive cluster satellites in the TNG-Cluster simulation: Existence and detectability}",
      journal = {\aap},
         year = 2024,
        month = jun,
       volume = {686},
          eid = {A86},
        pages = {A86},
          doi = {10.1051/0004-6361/202348583},
archivePrefix = {arXiv},
       eprint = {2311.06337},
 primaryClass = {astro-ph.GA},
       adsurl = {https://ui.adsabs.harvard.edu/abs/2024A&A...686A..86R}
}

@article{anders1989abundances,
  title={Abundances of the elements: Meteoritic and solar},
  author={Anders, Edward and Grevesse, Nicolas},
  journal={Geochimica et Cosmochimica acta},
  volume={53},
  number={1},
  pages={197--214},
  year={1989},
  publisher={Elsevier}
}

@article{nandra2013hot,
  title={The Hot and Energetic Universe: A White Paper presenting the science theme motivating the Athena+ mission},
  author={Nandra, Kirpal and Barret, Didier and Barcons, Xavier and Fabian, Andy and Herder, Jan-Willem den and Piro, Luigi and Watson, Mike and Adami, Christophe and Aird, James and Afonso, Jose Manuel and others},
  journal={arXiv preprint arXiv:1306.2307},
  year={2013}
}

@article{ponti2023abundance,
  title={Abundance and temperature of the outer hot circumgalactic medium-The SRG/eROSITA view of the soft X-ray background in the eFEDS field},
  author={Ponti, G and Zheng, X and Locatelli, N and Bianchi, S and Zhang, Y and Anastasopoulou, K and Comparat, J and Dennerl, K and Freyberg, M and Haberl, F and others},
  journal={A\&A},
  volume={674},
  pages={A195},
  year={2023},
  publisher={EDP Sciences}
}

@article{locatelli2024warm,
  title={The warm-hot circumgalactic medium of the Milky Way as seen by eROSITA},
  author={Locatelli, N and Ponti, G and Zheng, X and Merloni, A and Becker, W and Comparat, J and Dennerl, K and Freyberg, MJ and Sasaki, M and Yeung, MCH},
  journal={A\&A},
  volume={681},
  pages={A78},
  year={2024},
  publisher={EDP Sciences}
}

@article{koutroumpa2007ovii,
  title={OVII and OVIII line emission in the diffuse soft X-ray background: heliospheric and galactic contributions},
  author={Koutroumpa, Dimitra and Acero, F and Lallement, Rosine and Ballet, J and Kharchenko, V},
  journal={A\&A},
  volume={475},
  number={3},
  pages={901--914},
  year={2007},
  publisher={EDP Sciences}
}

@article{cui2020hubs,
  title={HUBS: hot universe baryon surveyor},
  author={Cui, W and Chen, L-B and Gao, B and Guo, F-L and Jin, H and Wang, G-L and Wang, L and Wang, J-J and Wang, W and Wang, Z-S and others},
  journal={JLTP},
  volume={199},
  pages={502--509},
  year={2020},
  publisher={Springer}
}

@article{weng2024physical,
  title={The physical origins of gas in the circumgalactic medium using observationally motivated TNG50 mocks},
  author={Weng, Simon and P{\'e}roux, C{\'e}line and Ramesh, Rahul and Nelson, Dylan and Sadler, Elaine M and Zwaan, Martin and Bollo, Victoria and Casavecchia, Benedetta},
  journal={MNRAS},
  volume={527},
  number={2},
  pages={3494--3516},
  year={2024},
  publisher={Oxford University Press},
url={https://academic.oup.com/mnras/article/527/2/3494/7371665}
}

@article{dave2019simba,
  title={SIMBA: Cosmological simulations with black hole growth and feedback},
  author={Dav{\'e}, Romeel and Angl{\'e}s-Alc{\'a}zar, Daniel and Narayanan, Desika and Li, Qi and Rafieferantsoa, Mika H and Appleby, Sarah},
  journal={MNRAS},
  volume={486},
  number={2},
  pages={2827--2849},
  year={2019},
  publisher={Oxford University Press}
}

@article{shreeram2025quantifying,
  title={Quantifying observational projection effects with a simulation-based hot CGM model},
  author={Shreeram, Soumya and Comparat, Johan and Merloni, Andrea and Zhang, Yi and Ponti, Gabriele and Nandra, Kirpal and ZuHone, John and Marini, Ilaria and Vladutescu-Zopp, Stephan and Popesso, Paola and others},
  journal={A\&A},
  volume={697},
  pages={A22},
  year={2025},
  publisher={EDP Sciences}
}

@ARTICLE{Comparat2019agnmodel,
       author = {{Comparat}, J. and {Merloni}, A. and {Salvato}, M. and {Nandra}, K. and {Boller}, T. and {Georgakakis}, A. and {Finoguenov}, A. and {Dwelly}, T. and {Buchner}, J. and {Del Moro}, A. and {Clerc}, N. and {Wang}, Y. and {Zhao}, G. and {Prada}, F. and {Yepes}, G. and {Brusa}, M. and {Krumpe}, M. and {Liu}, T.},
        title = "{Active galactic nuclei and their large-scale structure: an eROSITA mock catalogue}",
      journal = {\mnras},
         year = 2019,
        month = aug,
       volume = {487},
       number = {2},
        pages = {2005-2029},
          doi = {10.1093/mnras/stz1390},
archivePrefix = {arXiv},
       eprint = {1901.10866},
 primaryClass = {astro-ph.GA},
       adsurl = {https://ui.adsabs.harvard.edu/abs/2019MNRAS.487.2005C}
}

@ARTICLE{Habouzit2019xlf,
       author = {{Habouzit}, M{\'e}lanie and {Genel}, Shy and {Somerville}, Rachel S. and {Kocevski}, Dale and {Hirschmann}, Michaela and {Dekel}, Avishai and {Choi}, Ena and {Nelson}, Dylan and {Pillepich}, Annalisa and {Torrey}, Paul and {Hernquist}, Lars and {Vogelsberger}, Mark and {Weinberger}, Rainer and {Springel}, Volker},
        title = "{Linking galaxy structural properties and star formation activity to black hole activity with IllustrisTNG}",
      journal = {\mnras},
         year = 2019,
        month = apr,
       volume = {484},
       number = {4},
        pages = {4413-4443},
          doi = {10.1093/mnras/stz102},
archivePrefix = {arXiv},
       eprint = {1809.05588},
 primaryClass = {astro-ph.GA},
       adsurl = {https://ui.adsabs.harvard.edu/abs/2019MNRAS.484.4413H}
}

@article{sijacki2015illustris,
  title={The Illustris simulation: the evolving population of black holes across cosmic time},
  author={Sijacki, Debora and Vogelsberger, Mark and Genel, Shy and Springel, Volker and Torrey, Paul and Snyder, Gregory F and Nelson, Dylan and Hernquist, Lars},
  journal={MNRAS},
  volume={452},
  number={1},
  pages={575--596},
  year={2015},
  publisher={Oxford University Press}
}

@article{volonteri2016cosmic,
  title={The cosmic evolution of massive black holes in the Horizon-AGN simulation},
  author={Volonteri, Marta and Dubois, Yohan and Pichon, Christophe and Devriendt, Julien},
  journal={MNRAS},
  volume={460},
  number={3},
  pages={2979--2996},
  year={2016},
  publisher={Oxford University Press}
}

@article{rosas2016supermassive,
  title={Supermassive black holes in the EAGLE Universe. Revealing the observables of their growth},
  author={Rosas-Guevara, Yetli and Bower, Richard G and Schaye, Joop and McAlpine, Stuart and Dalla Vecchia, Claudio and Frenk, Carlos S and Schaller, Matthieu and Theuns, Tom},
  journal={MNRAS},
  volume={462},
  number={1},
  pages={190--205},
  year={2016},
  publisher={The Royal Astronomical Society}
}

@article{aird2015evolution,
  title={The evolution of the X-ray luminosity functions of unabsorbed and absorbed AGNs out to z~ 5},
  author={Aird, James and Coil, Alison L and Georgakakis, Antonis and Nandra, Kirpal and Barro, Guillermo and P{\'e}rez-Gonz{\'a}lez, Pablo G},
  journal={MNRAS},
  volume={451},
  number={2},
  pages={1892--1927},
  year={2015},
  publisher={Oxford University Press},
url={https://academic.oup.com/mnras/article/451/2/1892/1747784}
}

@ARTICLE{aird2017x,
       author = {{Aird}, J. and {Coil}, A.~L. and {Georgakakis}, A.},
        title = "{X-rays across the galaxy population - I. Tracing the main sequence of star formation}",
      journal = {\mnras},
         year = 2017,
        month = mar,
       volume = {465},
       number = {3},
        pages = {3390-3415},
          doi = {10.1093/mnras/stw2932},
archivePrefix = {arXiv},
       eprint = {1611.03508},
 primaryClass = {astro-ph.GA},
       adsurl = {https://ui.adsabs.harvard.edu/abs/2017MNRAS.465.3390A}
}

@article{lehmer2016evolution,
  title={The evolution of normal galaxy X-ray emission through cosmic history: constraints from the 6 Ms Chandra Deep Field-South},
  author={Lehmer, BD and Basu-Zych, AR and Mineo, S and Brandt, WN and Eufrasio, RT and Fragos, T and Hornschemeier, AE and Luo, B and Xue, YQ and Bauer, Franz Erik and others},
  journal={ApJ},
  volume={825},
  number={1},
  pages={7},
  year={2016},
  publisher={IOP Publishing}
}

@article{brandt2015cosmic,
  title={Cosmic X-ray surveys of distant active galaxies: The demographics, physics, and ecology of growing supermassive black holes},
  author={Brandt, WN and Alexander, DM},
  journal={ARA\&A},
  volume={23},
  number={1},
  pages={1},
  year={2015},
  publisher={Springer}
}

@article{foreman2013emcee,
  title={emcee: the MCMC hammer},
  author={Foreman-Mackey, Daniel and Hogg, David W and Lang, Dustin and Goodman, Jonathan},
  journal={PASP},
  volume={125},
  number={925},
  pages={306},
  year={2013},
  publisher={IOP Publishing}
}

@article{hastings1970monte,
  title={Monte Carlo sampling methods using Markov chains and their applications},
  author={Hastings, W Keith},
  year={1970},
  publisher={Oxford University Press}
}

@article{donahue2022baryon,
  title={Baryon cycles in the biggest galaxies},
  author={Donahue, Megan and Voit, G Mark},
  journal={Physics Reports},
  volume={973},
  pages={1--109},
  year={2022},
  url={https://www.sciencedirect.com/science/article/pii/S0370157322001302?via%3Dihub#b76},
  publisher={Elsevier}
}

@article{crain2023hydrodynamical,
  title={Hydrodynamical simulations of the galaxy population: enduring successes and outstanding challenges},
  author={Crain, Robert A and van de Voort, Freeke},
  journal={ARAA},
  volume={61},
  number={1},
  pages={473--515},
  year={2023},
  publisher={Annual Reviews}
}

@article{zhang2024hot2,
  title={The hot circumgalactic medium in the eROSITA All-Sky Survey-II. Scaling relations between X-ray luminosity and galaxies’ mass},
  author={Zhang, Yi and Comparat, Johan and Ponti, Gabriele and Merloni, Andrea and Nandra, Kirpal and Haberl, Frank and Truong, Nhut and Pillepich, Annalisa and Locatelli, Nicola and Zhang, Xiaoyuan and others},
  journal={A\&A},
  volume={690},
  pages={A268},
  year={2024},
  publisher={EDP Sciences}
}

@article{fabbiano2006populations,
  title={Populations of X-ray sources in galaxies},
  author={Fabbiano, G},
  journal={Annu. Rev. Astron. Astrophys.},
  volume={44},
  number={1},
  pages={323--366},
  year={2006},
  publisher={Annual Reviews}
}

@ARTICLE{Grimm2003hmxb,
       author = {{Grimm}, H. -J. and {Gilfanov}, M. and {Sunyaev}, R.},
        title = "{High-mass X-ray binaries as a star formation rate indicator in distant galaxies}",
      journal = {\mnras},
         year = 2003,
        month = mar,
       volume = {339},
       number = {3},
        pages = {793-809},
          doi = {10.1046/j.1365-8711.2003.06224.x},
archivePrefix = {arXiv},
       eprint = {astro-ph/0205371},
 primaryClass = {astro-ph},
       adsurl = {https://ui.adsabs.harvard.edu/abs/2003MNRAS.339..793G}
}

@ARTICLE{Shtykovskiy2005smc,
       author = {{Shtykovskiy}, P. and {Gilfanov}, M.},
        title = "{High-mass X-ray binaries in the Small Magellanic Cloud: the luminosity function}",
      journal = {\mnras},
         year = 2005,
        month = sep,
       volume = {362},
       number = {3},
        pages = {879-890},
          doi = {10.1111/j.1365-2966.2005.09320.x},
archivePrefix = {arXiv},
       eprint = {astro-ph/0503477},
 primaryClass = {astro-ph},
       adsurl = {https://ui.adsabs.harvard.edu/abs/2005MNRAS.362..879S}
}

@article{mineo2012x,
  title={X-ray emission from star-forming galaxies--I. High-mass X-ray binaries},
  author={Mineo, S and Gilfanov, M and Sunyaev, R},
  journal={MNRAS},
  volume={419},
  number={3},
  pages={2095--2115},
  year={2012},
  publisher={The Royal Astronomical Society}
}

@article{gilfanov2004low,
  title={Low-mass X-ray binaries as a stellar mass indicator for the host galaxy},
  author={Gilfanov, Marat},
  journal={MNRAS},
  volume={349},
  number={1},
  pages={146--168},
  year={2004},
  publisher={Blackwell Science Ltd Oxford, UK}
}

@article{boroson2011revisiting,
  title={Revisiting with Chandra the scaling relations of the X-ray emission components (binaries, nuclei, and hot gas) of early-type galaxies},
  author={Boroson, Bram and Kim, Dong-Woo and Fabbiano, Giuseppina},
  journal={ApJ},
  volume={729},
  number={1},
  pages={12},
  year={2011},
  publisher={IOP Publishing}
}

@ARTICLE{Zhang2012lxrb,
       author = {{Zhang}, Z. and {Gilfanov}, M. and {Bogd{\'a}n}, {\'A}.},
        title = "{Dependence of the low-mass X-ray binary population on stellar age}",
      journal = {\aap},
         year = 2012,
        month = oct,
       volume = {546},
          eid = {A36},
        pages = {A36},
          doi = {10.1051/0004-6361/201219015},
archivePrefix = {arXiv},
       eprint = {1202.2331},
 primaryClass = {astro-ph.HE},
       adsurl = {https://ui.adsabs.harvard.edu/abs/2012A&A...546A..36Z}
}

@article{lehmer2019x,
  title={X-ray binary luminosity function scaling relations for local galaxies based on subgalactic modeling},
  author={Lehmer, Bret D and Eufrasio, Rafael T and Tzanavaris, Panayiotis and Basu-Zych, Antara and Fragos, Tassos and Prestwich, Andrea and Yukita, Mihoko and Zezas, Andreas and Hornschemeier, Ann E and Ptak, Andrew},
  journal={ApJ Supplement Series},
  volume={243},
  number={1},
  pages={3},
  year={2019},
  publisher={IOP Publishing}
}

@article{ishiyama2021uchuu,
  title={The Uchuu simulations: Data Release 1 and dark matter halo concentrations},
  author={Ishiyama, Tomoaki and Prada, Francisco and Klypin, Anatoly A and Sinha, Manodeep and Metcalf, R Benton and Jullo, Eric and Altieri, Bruno and Cora, Sof{\'\i}a A and Croton, Darren and de La Torre, Sylvain and others},
  journal={MNRAS},
  volume={506},
  number={3},
  pages={4210--4231},
  year={2021},
  publisher={Oxford University Press}
}

@article{behroozi2019universemachine,
  title={UniverseMachine: The correlation between galaxy growth and dark matter halo assembly from z= 0- 10},
  author={Behroozi, Peter and Wechsler, Risa H and Hearin, Andrew P and Conroy, Charlie},
  journal={MNRAS},
  volume={488},
  number={3},
  pages={3143--3194},
  year={2019},
  publisher={Oxford University Press}
}

@article{donnari2019star,
  title={The star formation activity of IllustrisTNG galaxies: main sequence, UVJ diagram, quenched fractions, and systematics},
  author={Donnari, Martina and Pillepich, Annalisa and Nelson, Dylan and Vogelsberger, Mark and Genel, Shy and Weinberger, Rainer and Marinacci, Federico and Springel, Volker and Hernquist, Lars},
  journal={MNRAS},
  volume={485},
  number={4},
  pages={4817--4840},
  year={2019},
  publisher={Oxford University Press}
}

@article{donnari2021quenched,
  title={Quenched fractions in the IllustrisTNG simulations: comparison with observations and other theoretical models},
  author={Donnari, Martina and Pillepich, Annalisa and Nelson, Dylan and Marinacci, Federico and Vogelsberger, Mark and Hernquist, Lars},
  journal={MNRAS},
  volume={506},
  number={4},
  pages={4760--4780},
  year={2021},
  url={https://academic.oup.com/mnras/article/506/4/4760/6318380#283964124},  
  publisher={Oxford University Press}
}

@article{aird2013primus,
  title={PRIMUS: AN OBSERVATIONALLY MOTIVATED MODEL TO CONNECT THE EVOLUTION OF THE ACTIVE GALACTIC NUCLEUS AND GALAXY POPULATIONS OUT TO z~ 1},
  author={Aird, James and Coil, Alison L and Moustakas, John and Diamond-Stanic, Aleksandar M and Blanton, Michael R and Cool, Richard J and Eisenstein, Daniel J and Wong, Kenneth C and Zhu, Guangtun},
  journal={ApJ},
  volume={775},
  number={1},
  pages={41},
  year={2013},
  publisher={IOP Publishing}
}

@article{hasinger2008absorption,
  title={Absorption properties and evolution of active galactic nuclei},
  author={Hasinger, G},
  journal={A\&A},
  volume={490},
  number={3},
  pages={905--922},
  year={2008},
  publisher={EDP Sciences}
}

@article{marconi2003relation,
  title={The relation between black hole mass, bulge mass, and near-infrared luminosity},
  author={Marconi, Alessandro and Hunt, Leslie K},
  journal={ApJ},
  volume={589},
  number={1},
  pages={L21},
  year={2003},
  publisher={IOP Publishing}
}

@article{lau2024x,
  title={X-Raying CAMELS: Constraining Baryonic Feedback in the Circumgalactic Medium with the CAMELS Simulations and eRASS X-Ray Observations},
  author={Lau, Erwin T and Nagai, Daisuke and Bogd{\'a}n, {\'A}kos and Medlock, Isabel and Oppenheimer, Benjamin D and Battaglia, Nicholas and Angl{\'e}s-Alc{\'a}zar, Daniel and Genel, Shy and Ni, Yueying and Villaescusa-Navarro, Francisco},
  journal={ApJ},
  volume={984},
  number={2},
  pages={190},
  year={2025},
  publisher={IOP Publishing}
}

@ARTICLE{Ricci2017rad,
       author = {{Ricci}, Claudio and {Trakhtenbrot}, Benny and {Koss}, Michael J. and {Ueda}, Yoshihiro and {Schawinski}, Kevin and {Oh}, Kyuseok and {Lamperti}, Isabella and {Mushotzky}, Richard and {Treister}, Ezequiel and {Ho}, Luis C. and {Weigel}, Anna and {Bauer}, Franz E. and {Paltani}, Stephane and {Fabian}, Andrew C. and {Xie}, Yanxia and {Gehrels}, Neil},
        title = "{The close environments of accreting massive black holes are shaped by radiative feedback}",
      journal = {\nat},
         year = 2017,
        month = sep,
       volume = {549},
       number = {7673},
        pages = {488-491},
          doi = {10.1038/nature23906},
archivePrefix = {arXiv},
       eprint = {1709.09651},
 primaryClass = {astro-ph.HE},
       adsurl = {https://ui.adsabs.harvard.edu/abs/2017Natur.549..488R}
}

@article{buchner2017galaxy,
  title={Galaxy gas as obscurer--II. Separating the galaxy-scale and nuclear obscurers of active galactic nuclei},
  author={Buchner, Johannes and Bauer, Franz E},
  journal={MNRAS},
  volume={465},
  number={4},
  pages={4348--4362},
  year={2017},
  publisher={Oxford University Press}
}

@article{ueda2014toward,
  title={Toward the standard population synthesis model of the x-ray background: Evolution of X-ray luminosity and absorption functions of active galactic nuclei including Compton-thick populations},
  author={Ueda, Yoshihiro and Akiyama, Masayuki and Hasinger, G{\"u}nther and Miyaji, Takamitsu and Watson, Michael G},
  journal={ApJ},
  volume={786},
  number={2},
  pages={104},
  year={2014},
  publisher={IOP Publishing}
}

@article{buchner2015obscuration,
  title={Obscuration-dependent evolution of active galactic nuclei},
  author={Buchner, Johannes and Georgakakis, Antonis and Nandra, Kirpal and Brightman, Murray and Menzel, Marie-Luise and Liu, Zhu and Hsu, Li-Ting and Salvato, Mara and Rangel, Cyprian and Aird, James and others},
  journal={ApJ},
  volume={802},
  number={2},
  pages={89},
  year={2015},
  publisher={IOP Publishing}
}

@ARTICLE{popesso2024hot,
       author = {{Popesso}, P. and {Biviano}, A. and {Marini}, I. and {Dolag}, K. and {Vladutescu-Zopp}, S. and {Csizi}, B. and {Biffi}, V. and {Lamer}, G. and {Robothan}, A. and {Bravo}, M. and {Lovisari}, L. and {Ettori}, S. and {Angelinelli}, M. and {Driver}, S. and {Toptun}, V. and {Dev}, A. and {Mazengo}, D. and {Merloni}, A. and {Comparat}, J. and {Ponti}, G. and {Mroczkowski}, T. and {Bulbul}, E. and {Grandis}, S. and {Bahar}, E.},
        title = "{The hot gas mass fraction in halos. From Milky Way-like groups to massive clusters}",
      journal = {arXiv e-prints},
         year = 2024,
        month = nov,
          eid = {arXiv:2411.16555},
        pages = {arXiv:2411.16555},
          doi = {10.48550/arXiv.2411.16555},
archivePrefix = {arXiv},
       eprint = {2411.16555},
 primaryClass = {astro-ph.GA},
       adsurl = {https://ui.adsabs.harvard.edu/abs/2024arXiv241116555P}
}

@ARTICLE{popesso2024average,
       author = {{Popesso}, P. and {Marini}, I. and {Dolag}, K. and {Lamer}, G. and {Csizi}, B. and {Biffi}, V. and {Robothan}, A. and {Bravo}, M. and {Biviano}, A. and {Vladutesku-Zopp}, S. and {Lovisari}, L. and {Ettori}, S. and {Angelinelli}, M. and {Driver}, S. and {Toptun}, V. and {Dev}, A. and {Mazengo}, D. and {Merloni}, A. and {Zhang}, Y. and {Comparat}, J. and {Ponti}, G. and {Mroczkowski}, T. and {Bulbul}, E.},
        title = "{Average X-ray properties of galaxy groups. From Milky Way-like halos to massive clusters}",
      journal = {arXiv e-prints},
         year = 2024,
        month = nov,
          eid = {arXiv:2411.17120},
        pages = {arXiv:2411.17120},
          doi = {10.48550/arXiv.2411.17120},
archivePrefix = {arXiv},
       eprint = {2411.17120},
 primaryClass = {astro-ph.GA},
       adsurl = {https://ui.adsabs.harvard.edu/abs/2024arXiv241117120P}
}

@ARTICLE{moster2020emerge,
       author = {{Moster}, Benjamin P. and {Naab}, Thorsten and {White}, Simon D.~M.},
        title = "{EMERGE - empirical constraints on the formation of passive galaxies}",
      journal = {\mnras},
         year = 2020,
        month = dec,
       volume = {499},
       number = {4},
        pages = {4748-4767},
          doi = {10.1093/mnras/staa3019},
archivePrefix = {arXiv},
       eprint = {1910.09552},
 primaryClass = {astro-ph.GA},
       adsurl = {https://ui.adsabs.harvard.edu/abs/2020MNRAS.499.4748M}
}

@article{duras2020universal,
  title={Universal bolometric corrections for active galactic nuclei over seven luminosity decades},
  author={Duras, F and Bongiorno, A and Ricci, F and Piconcelli, E and Shankar, F and Lusso, E and Bianchi, S and Fiore, F and Maiolino, R and Marconi, A and others},
  journal={A\&A},
  volume={636},
  pages={A73},
  year={2020},
  publisher={EDP Sciences}
}

@article{collin2002quasars,
  title={Are quasars accreting at super-Eddington rates?},
  author={Collin, Suzy and Boisson, Catherine and Mouchet, Martine and Dumont, A-M and Coup{\'e}, S{\'e}verine and Porquet, Delphine and Rokaki, Evlabia},
  journal={A\&A},
  volume={388},
  number={3},
  pages={771--786},
  year={2002},
  publisher={EDP Sciences}
}

@article{buchner2024genuine,
  title={Genuine Retrieval of the AGN Host Stellar Population (GRAHSP)},
  author={Buchner, Johannes and Starck, Hattie and Salvato, Mara and Netzer, Hagai and Igo, Zsofi and Laloux, Brivael and Georgakakis, Antonis and Gauger, Isabelle and Olechowska, Anna and Lopez, Nicolas and others},
  journal={A\&A},
  volume={692},
  pages={A161},
  year={2024},
  publisher={EDP Sciences}
}

@ARTICLE{Zheng2024ovii,
       author = {{Zheng}, Xueying and {Ponti}, Gabriele and {Locatelli}, Nicola and {Sanders}, Jeremy and {Merloni}, Andrea and {Becker}, Werner and {Comparat}, Johan and {Dennerl}, Konrad and {Freyberg}, Michael and {Maitra}, Chandreyee and {Sasaki}, Manami and {Strong}, Andrew and {Yeung}, Michael C.~H.},
        title = "{eROSITA narrowband maps at the energies of soft X-ray emission lines}",
      journal = {\aap},
         year = 2024,
        month = sep,
       volume = {689},
          eid = {A328},
        pages = {A328},
          doi = {10.1051/0004-6361/202449398},
archivePrefix = {arXiv},
       eprint = {2401.17310},
 primaryClass = {astro-ph.GA},
       adsurl = {https://ui.adsabs.harvard.edu/abs/2024A&A...689A.328Z}
}

@INPROCEEDINGS{mushotzky2019advanced,
       author = {{Mushotzky}, Richard and {Aird}, James and {Barger}, Amy J. and {Cappelluti}, Nico and {Chartas}, George and {Corrales}, L{\'\i}a and {Eufrasio}, Rafael and {Fabian}, Andrew C. and {Falcone}, Abraham D. and {Gallo}, Elena and {Gilli}, Roberto and {Grant}, Catherine E. and {Hardcastle}, Martin and {Hodges-Kluck}, Edmund and {Kara}, Erin and {Koss}, Michael and {Li}, Hui and {Lisse}, Carey M. and {Loewenstein}, Michael and {Markevitch}, Maxim and {Meyer}, Eileen T. and {Miller}, Eric D. and {Mulchaey}, John and {Petre}, Robert and {Ptak}, Andrew J. and {Reynolds}, Christopher S. and {Russell}, Helen R. and {Safi-Harb}, Samar and {Smith}, Randall K. and {Snios}, Bradford and {Tombesi}, Francesco and {Valencic}, Lynne and {Walker}, Stephen A. and {Williams}, Brian J. and {Winter}, Lisa M. and {Yamaguchi}, Hiroya and {Zhang}, William W. and {Arenberg}, Jon and {Brandt}, Niel and {Burrows}, David N. and {Georganopoulos}, Markos and {Miller}, Jon M. and {Norman}, Colin A. and {Rosati}, Piero},
        title = "{The Advanced X-ray Imaging Satellite}",
    booktitle = {BAAS},
         year = 2019,
       volume = {51},
        month = sep,
          eid = {107},
        pages = {107},
          doi = {10.48550/arXiv.1903.04083},
archivePrefix = {arXiv},
       eprint = {1903.04083},
 primaryClass = {astro-ph.HE},
       adsurl = {https://ui.adsabs.harvard.edu/abs/2019BAAS...51g.107M}
}

@article{schaye2023flamingo,
  title={The FLAMINGO project: cosmological hydrodynamical simulations for large-scale structure and galaxy cluster surveys},
  author={Schaye, Joop and Kugel, Roi and Schaller, Matthieu and Helly, John C and Braspenning, Joey and Elbers, Willem and McCarthy, Ian G and van Daalen, Marcel P and Vandenbroucke, Bert and Frenk, Carlos S and others},
  journal={MNRAS},
  volume={526},
  number={4},
  pages={4978--5020},
  year={2023},
  publisher={Oxford University Press}
}

@ARTICLE{Crain2015eagle,
       author = {{Crain}, Robert A. and {Schaye}, Joop and {Bower}, Richard G. and {Furlong}, Michelle and {Schaller}, Matthieu and {Theuns}, Tom and {Dalla Vecchia}, Claudio and {Frenk}, Carlos S. and {McCarthy}, Ian G. and {Helly}, John C. and {Jenkins}, Adrian and {Rosas-Guevara}, Yetli M. and {White}, Simon D.~M. and {Trayford}, James W.},
        title = "{The EAGLE simulations of galaxy formation: calibration of subgrid physics and model variations}",
      journal = {\mnras},
         year = 2015,
        month = jun,
       volume = {450},
       number = {2},
        pages = {1937-1961},
          doi = {10.1093/mnras/stv725},
archivePrefix = {arXiv},
       eprint = {1501.01311},
 primaryClass = {astro-ph.GA},
       adsurl = {https://ui.adsabs.harvard.edu/abs/2015MNRAS.450.1937C}
}

@ARTICLE{Schaye2015eagle,
       author = {{Schaye}, Joop and {Crain}, Robert A. and {Bower}, Richard G. and {Furlong}, Michelle and {Schaller}, Matthieu and {Theuns}, Tom and {Dalla Vecchia}, Claudio and {Frenk}, Carlos S. and {McCarthy}, I.~G. and {Helly}, John C. and {Jenkins}, Adrian and {Rosas-Guevara}, Y.~M. and {White}, Simon D.~M. and {Baes}, Maarten and {Booth}, C.~M. and {Camps}, Peter and {Navarro}, Julio F. and {Qu}, Yan and {Rahmati}, Alireza and {Sawala}, Till and {Thomas}, Peter A. and {Trayford}, James},
        title = "{The EAGLE project: simulating the evolution and assembly of galaxies and their environments}",
      journal = {\mnras},
         year = 2015,
        month = jan,
       volume = {446},
       number = {1},
        pages = {521-554},
          doi = {10.1093/mnras/stu2058},
archivePrefix = {arXiv},
       eprint = {1407.7040},
 primaryClass = {astro-ph.GA},
       adsurl = {https://ui.adsabs.harvard.edu/abs/2015MNRAS.446..521S}
}

@article{medlock2025quantifying, title={Quantifying Baryonic Feedback on the Warm--Hot Circumgalactic Medium in CAMELS Simulations}, author={Medlock, Isabel and Neufeld, Chloe and Nagai, Daisuke and Angl{'e}s-Alc{'a}zar, Daniel and Genel, Shy and Oppenheimer, Benjamin D and Sims, Xavier and Singh, Priyanka and Villaescusa-Navarro, Francisco}, journal={ApJ}, volume={980}, number={1}, pages={61}, year={2025}, publisher={IOP Publishing} }

@ARTICLE{Beck2016magneticum,
       author = {{Beck}, A.~M. and {Murante}, G. and {Arth}, A. and {Remus}, R. -S. and {Teklu}, A.~F. and {Donnert}, J.~M.~F. and {Planelles}, S. and {Beck}, M.~C. and {F{\"o}rster}, P. and {Imgrund}, M. and {Dolag}, K. and {Borgani}, S.},
        title = "{An improved SPH scheme for cosmological simulations}",
      journal = {\mnras},
         year = 2016,
        month = jan,
       volume = {455},
       number = {2},
        pages = {2110-2130},
          doi = {10.1093/mnras/stv2443},
archivePrefix = {arXiv},
       eprint = {1502.07358},
 primaryClass = {astro-ph.CO},
       adsurl = {https://ui.adsabs.harvard.edu/abs/2016MNRAS.455.2110B}
}

@ARTICLE{Dolag2005magneticum,
       author = {{Dolag}, Klaus and {Grasso}, Dario and {Springel}, Volker and {Tkachev}, Igor},
        title = "{Constrained simulations of the magnetic field in the local Universe and the propagation of ultrahigh energy cosmic rays}",
      journal = {\jcap},
         year = 2005,
        month = jan,
       volume = {2005},
       number = {1},
          eid = {009},
        pages = {009},
          doi = {10.1088/1475-7516/2005/01/009},
archivePrefix = {arXiv},
       eprint = {astro-ph/0410419},
 primaryClass = {astro-ph},
       adsurl = {https://ui.adsabs.harvard.edu/abs/2005JCAP...01..009D}
}

@ARTICLE{Taylor2020z_0_18,
       author = {{Taylor}, Edward N. and {Cluver}, Michelle E. and {Duffy}, Alan and {Gurri}, Pol and {Hoekstra}, Henk and {Sonnenfeld}, Alessandro and {Bremer}, Malcolm N. and {Brouwer}, Margot M. and {Chisari}, Nora Elisa and {Dvornik}, Andrej and {Erben}, Thomas and {Hildebrandt}, Hendrik and {Hopkins}, Andrew M. and {Kelvin}, Lee S. and {Phillipps}, Steven and {Robotham}, Aaron S.~G. and {Sif{\'o}n}, Cristob{\'a}l and {Vakili}, Mohammadjavad and {Wright}, Angus H.},
        title = "{GAMA + KiDS: empirical correlations between halo mass and other galaxy properties near the knee of the stellar-to-halo mass relation}",
      journal = {\mnras},
         year = 2020,
        month = dec,
       volume = {499},
       number = {2},
        pages = {2896-2911},
          doi = {10.1093/mnras/staa2648},
archivePrefix = {arXiv},
       eprint = {2006.10040},
 primaryClass = {astro-ph.GA},
       adsurl = {https://ui.adsabs.harvard.edu/abs/2020MNRAS.499.2896T}
}

@ARTICLE{Leauthaud2012z_0_2,
       author = {{Leauthaud}, Alexie and {Tinker}, Jeremy and {Bundy}, Kevin and {Behroozi}, Peter S. and {Massey}, Richard and {Rhodes}, Jason and {George}, Matthew R. and {Kneib}, Jean-Paul and {Benson}, Andrew and {Wechsler}, Risa H. and {Busha}, Michael T. and {Capak}, Peter and {Cort{\^e}s}, Marina and {Ilbert}, Olivier and {Koekemoer}, Anton M. and {Le F{\`e}vre}, Oliver and {Lilly}, Simon and {McCracken}, Henry J. and {Salvato}, Mara and {Schrabback}, Tim and {Scoville}, Nick and {Smith}, Tristan and {Taylor}, James E.},
        title = "{New Constraints on the Evolution of the Stellar-to-dark Matter Connection: A Combined Analysis of Galaxy-Galaxy Lensing, Clustering, and Stellar Mass Functions from z = 0.2 to z =1}",
      journal = {\apj},
         year = 2012,
        month = jan,
       volume = {744},
       number = {2},
          eid = {159},
        pages = {159},
          doi = {10.1088/0004-637X/744/2/159},
archivePrefix = {arXiv},
       eprint = {1104.0928},
 primaryClass = {astro-ph.CO},
       adsurl = {https://ui.adsabs.harvard.edu/abs/2012ApJ...744..159L}
}

@ARTICLE{Coupon2015_z_0_8,
       author = {{Coupon}, J. and {Arnouts}, S. and {van Waerbeke}, L. and {Moutard}, T. and {Ilbert}, O. and {van Uitert}, E. and {Erben}, T. and {Garilli}, B. and {Guzzo}, L. and {Heymans}, C. and {Hildebrandt}, H. and {Hoekstra}, H. and {Kilbinger}, M. and {Kitching}, T. and {Mellier}, Y. and {Miller}, L. and {Scodeggio}, M. and {Bonnett}, C. and {Branchini}, E. and {Davidzon}, I. and {De Lucia}, G. and {Fritz}, A. and {Fu}, L. and {Hudelot}, P. and {Hudson}, M.~J. and {Kuijken}, K. and {Leauthaud}, A. and {Le F{\`e}vre}, O. and {McCracken}, H.~J. and {Moscardini}, L. and {Rowe}, B.~T.~P. and {Schrabback}, T. and {Semboloni}, E. and {Velander}, M.},
        title = "{The galaxy-halo connection from a joint lensing, clustering and abundance analysis in the CFHTLenS/VIPERS field}",
      journal = {\mnras},
         year = 2015,
        month = may,
       volume = {449},
       number = {2},
        pages = {1352-1379},
          doi = {10.1093/mnras/stv276},
archivePrefix = {arXiv},
       eprint = {1502.02867},
 primaryClass = {astro-ph.CO},
       adsurl = {https://ui.adsabs.harvard.edu/abs/2015MNRAS.449.1352C}
}

@ARTICLE{Girelli2020shmr,
       author = {{Girelli}, G. and {Pozzetti}, L. and {Bolzonella}, M. and {Giocoli}, C. and {Marulli}, F. and {Baldi}, M.},
        title = "{The stellar-to-halo mass relation over the past 12 Gyr. I. Standard {\ensuremath{\Lambda}}CDM model}",
      journal = {\aap},
         year = 2020,
        month = feb,
       volume = {634},
          eid = {A135},
        pages = {A135},
          doi = {10.1051/0004-6361/201936329},
archivePrefix = {arXiv},
       eprint = {2001.02230},
 primaryClass = {astro-ph.CO},
       adsurl = {https://ui.adsabs.harvard.edu/abs/2020A&A...634A.135G}
}

@article{alexander2012drives,
  title={What drives the growth of black holes?},
  author={Alexander, David M and Hickox, Ryan C},
  journal={New Astronomy Reviews},
  volume={56},
  number={4},
  pages={93--121},
  year={2012},
  publisher={Elsevier}
}

@ARTICLE{Comparat2025cross,
       author = {{Comparat}, Johan and {Merloni}, Andrea and {Ponti}, Gabriele and {Shreeram}, Soumya and {Zhang}, Yi and {Reiprich}, Thomas H. and {Liu}, Ang and {Seppi}, Riccardo and {Zhang}, Xiaoyuan and {Clerc}, Nicolas and {Nicola}, Andrina and {Nandra}, Kirpal and {Salvato}, Mara and {Malavasi}, Nicola},
        title = "{Cross-correlation between soft X-rays and galaxies: A new benchmark for galaxy evolution models}",
      journal = {\aap},
         year = 2025,
        month = may,
       volume = {697},
          eid = {A173},
        pages = {A173},
          doi = {10.1051/0004-6361/202554208},
archivePrefix = {arXiv},
       eprint = {2503.19796},
 primaryClass = {astro-ph.GA},
       adsurl = {https://ui.adsabs.harvard.edu/abs/2025A&A...697A.173C}
}

@article{zhang2018emission,
  title={Emission from the ionized gaseous halos of low-redshift galaxies and their neighbors},
  author={Zhang, Huanian and Zaritsky, Dennis and Behroozi, Peter},
  journal={ApJ},
  volume={861},
  number={1},
  pages={34},
  year={2018},
  publisher={IOP Publishing}
}

@article{wright2024baryon,
  title={The baryon cycle in modern cosmological hydrodynamical simulations},
  author={Wright, Ruby J and Somerville, Rachel S and Lagos, Claudia del P and Schaller, Matthieu and Dav{\'e}, Romeel and Angl{\'e}s-Alc{\'a}zar, Daniel and Genel, Shy},
  journal={MNRAS},
  volume={532},
  number={3},
  pages={3417--3440},
  year={2024},
  publisher={Oxford University Press}
}

\end{document}